\documentclass[twocolumn,preprintnumbers]{aastex6} 
\usepackage{afterpage}
\usepackage{capt-of}
\usepackage{diagbox}

\usepackage[figuresright]{rotating}
\usepackage{amsmath}
\usepackage{amssymb,amsmath,latexsym,graphics, graphicx,epsfig,multirow,comment,hyperref,appendix,feyn,slashed,xcolor,afterpage, makecell} 
\usepackage{booktabs}
\usepackage{tabularx}
\usepackage{wasysym}

\newcommand{\FeH}{\text{[Fe/H]} }

\newcommand {\be} {\begin {equation}}
\newcommand {\ee} {\end {equation}} 

\newcommand {\bes} {\begin {equation*}}
\newcommand {\ees} {\end {equation*}}

\newcolumntype{L}[1]{>{\raggedright\let\newline\\\arraybackslash\hspace{0pt}}m{#1}}
\newcolumntype{C}[1]{>{\centering\let\newline\\\arraybackslash\hspace{0pt}}m{#1}}
\newcolumntype{R}[1]{>{\raggedleft\let\newline\\\arraybackslash\hspace{0pt}}m{#1}}

\usepackage{csvsimple}

\makeatletter
\newcommand\footnoteref[1]{\protected@xdef\@thefnmark{\ref{#1}}\@footnotemark}
\makeatother

\DeclareRobustCommand{\Sec}[1]{Sec.~\ref{#1}}

\DeclareRobustCommand{\Tab}[1]{Table~\ref{#1}}

\DeclareRobustCommand{\Fig}[1]{Fig.~\ref{#1}}

\DeclareRobustCommand{\Eq}[1]{Eq.~(\ref{#1})}

\newcommand{\beq}{\begin{equation}}
\newcommand{\eeq}{\end{equation}}

\newcommand{\zacc}{z_\mathrm{acc}}

\newcommand{\mi}{\texttt{m12i}}
\newcommand{\mf}{\texttt{m12f}}

\begin{document}

\title{\vspace{-1cm}
Under the FIRElight: Stellar tracers of the local dark matter velocity distribution in the Milky Way
}

\author{Lina Necib}
\affil{Walter Burke Institute for Theoretical Physics,
California Institute of Technology, Pasadena, CA 91125, USA}

\author{\vspace{-0.4cm}Mariangela Lisanti}
\affil{Department of Physics, Princeton University, Princeton, NJ 08544, USA}

\author{\vspace{-0.4cm}Shea Garrison-Kimmel}
\affil{TAPIR, California Institute of Technology, Pasadena, CA 91125, USA}

\author{\vspace{-0.4cm}Andrew Wetzel}
\affil{Department of Physics, University of California, Davis, CA 95616, USA}

\author{\vspace{-0.4cm}Robyn Sanderson}
\affil{Department of Physics and Astronomy, University of Pennsylvania, Philadelphia, PA 19104, USA}
\affil{Center for Computational Astrophysics, Flatiron Institute, New York, NY 10010, USA}

\author{\vspace{-0.4cm}Philip F.\ Hopkins}
\affil{TAPIR, California Institute of Technology, Pasadena, CA 91125, USA}

\author{\vspace{-0.4cm}Claude-Andr\'e  Faucher-Gigu\`ere}
\affil{Department of Physics and Astronomy and CIERA, Northwestern University, Evanston, IL 60208, USA}

\author{\vspace{-0.4cm}Du\v{s}an Kere\v{s}}
\affil{Department of Physics, Center for Astrophysics and Space Sciences, University of California at San Diego, La Jolla, CA 92093, USA}

\begin{abstract}
The \emph{Gaia} era opens new possibilities for discovering the remnants of disrupted satellite galaxies in the Solar neighborhood.  
If the population of local accreted stars is correlated with the dark matter sourced by the same mergers, one can then map the dark matter distribution directly.  Using two cosmological zoom-in hydrodynamic simulations of Milky Way-mass galaxies from the \emph{Latte} suite of \textsc{Fire-2} simulations, we find a strong correlation between the velocity distribution of stars and dark matter at the solar circle that were accreted from luminous satellites. This correspondence holds for dark matter that is either relaxed or in kinematic substructure called debris flow, and is consistent between two simulated hosts with different merger histories.  The correspondence is more problematic for streams because of possible spatial offsets between the dark matter and stars.  We demonstrate how to reconstruct the dark matter velocity distribution from the observed properties of the accreted stellar population by properly accounting for the ratio of stars to dark matter contributed by individual mergers.  After demonstrating this method using the \textsc{Fire-2} simulations, we apply it to the Milky Way and use it to recover the dark matter velocity distribution associated with the recently discovered stellar debris field in the Solar neighborhood.  Based on results from \emph{Gaia}, we estimate that $42 ^{+26}_{-22}\%$ of the local dark matter that is accreted from luminous mergers is in debris flow.  
 \clearpage
\end{abstract}

\maketitle

\section{Introduction} 
\label{sec:intro}

In the $\Lambda$CDM paradigm, a dark matter (DM) host halo is built up hierarchically from galaxy mergers~\citep{1978MNRAS.183..341W,Diemand:2008in,Springel:2008cc,2011ApJ...740..102K}.  These  satellites also contribute stars, which may hold clues to the underlying DM distribution in the Milky Way.  In this work, we use simulations of Milky Way-mass galaxies from the \textsc{Feedback in Realistic Environments (Fire)}\footnote{\url{http://fire.northwestern.edu}} project~\citep{2017arXiv170206148H} to study the correlation between accreted stars and DM, and its dependence on galactic merger history.

The chemical abundance and phase-space distribution of an accreted stellar population can be used to infer properties of its parent galaxy~\citep{Helmi:2002iu, 2005ApJ...635..931B, Robertson:2005gv, Font:2005qs, DeLucia:2008gk, 2016ApJ...821....5D}.  In this fashion, \cite{2018arXiv180203414B} and \cite{ 2018arXiv180606038H} argued that the population of local accreted stars consists predominantly of debris from a disrupted satellite galaxy with original stellar mass $M_{\mathrm{*, total}} \sim 10^{7-8}$~M$_\odot$.  
This merger can potentially explain the observed density break in the halo at Galactocentric radii of $\sim 20$~kpc~\citep{2018ApJ...862L...1D}, as well as the population of globular clusters on highly radial orbits~\citep{Myeong:2018kfh}.  Referred to as the \emph{Gaia} Sausage or \emph{Gaia} Enceladus, this substructure comprises the majority of the local distribution of accreted stars (identified by both metallicity and kinematics), with the remaining fraction appearing to be nearly isotropic and metal poor.  

\cite{necib2018} showed that these findings have important implications for the local DM distribution, as they suggest that a non-trivial  fraction is in substructure.  This argument depends on whether stars that are tidally stripped from a satellite galaxy trace the DM that is removed from the same source.  The DM-stellar correspondence is not guaranteed for a variety of reasons.  First, stars are typically more tightly bound towards the center of a galaxy than DM, and thus have different initial phase-space structure. In an extreme case, a cuspy DM halo can admit a cored stellar distribution \citep{2013A&A...558L...3B}.  Additionally, the majority of stars are stripped only after the majority of DM because the latter is preferentially removed in the initial stages of satellite disruption.  Second, the mass-to-light ratio varies by orders of magnitude between galaxies \citep{2012AJ....144....4M}, so the relative mass of stars to DM that each contributes differs.  Therefore, even if one satellite contributes a significant fraction of accreted stars it may not contribute an equivalent fraction of the DM. These effects can be further exacerbated when restricting to a spatial volume like the solar neighborhood.  

In this work, we demonstrate how to reconstruct the properties of DM that is accreted from luminous satellites.  To organize the discussion, we classify the DM into three separate components that are delineated by relative accretion time.  The first component includes DM that was accreted at redshifts $z \gtrsim 3$ from the oldest mergers.  We refer to this component as `relaxed' in this work, though it has also been referred to as `virialized' in the literature.  
\cite{Herzog-Arbeitman:2017fte} demonstrated that this old DM population is well-traced by metal-poor stars using the \textsc{Eris} hydrodynamic simulation~\citep{Guedes:2011ux}.  In this case, convergence in the velocity distributions was reached for stars with iron abundance $\FeH \lesssim -3$.  This result motivated a first study using the RAVE-TGAS dataset to recover the velocity distribution of the local relaxed DM component~\citep{Herzog-Arbeitman:2017zbm}. 

We divide DM accreted from younger mergers into two separate categories: debris flow and streams.  Debris flow is an example of kinematic substructure that is spatially mixed on large scales.  It arises from the accretion of one or more older satellites that completed several orbital wraps~\citep{Lisanti:2011as,Kuhlen:2012fz}.  In this case, any structure in position-space is washed out, while velocity-space features are preserved~\citep{1999Natur.402...53H,2010MNRAS.408..935G}.  The properties of debris flow are quite similar between stars and DM, likely because the tidal debris is older and therefore more well-mixed~\citep{Lisanti:2014dva}.  These conclusions are based on studies of the \texttt{Via Lactea} DM-only simulation~\citep{Diemand:2008in} where star `particles' were painted onto the most bound DM `particles' in the satellite.  It should be repeated using a full hydrodynamic simulation, as we do here.

Streams, in contrast, are relics of the youngest mergers and are neither spatially  nor kinematically mixed.   They result from tidal debris that is torn off a satellite as it completes a small number of orbits~\citep{Zemp:2008gw, Vogelsberger:2008qb, Diemand:2008in,Kuhlen:2009vh,Maciejewski:2010gz,2011MNRAS.413.1419V, Elahi:2011dy}.  For these accretion events, the stars may not necessarily act as adequate tracers for the DM as has been noted in simulations of merging dwarf galaxies \citep{2008AN....329..934P} or of the Sagittarius stream~\citep{2012JCAP...08..027P}.

In this work, we study the correlation between stars and DM accreted from luminous satellites in two Milky-Way--mass halos with differing merger histories.  These two simulated galaxies share general properties of the Galactic disk and stellar halo~\citep{2018arXiv180610564S}, and are thus excellent systems in which to study the DM-stellar correlations of interest here.  
Our approach is to identify the stars and DM that originate from a given satellite galaxy and follow them as a function of time to see where they eventually end up relative to each other.  We find that stars from the oldest mergers trace the relaxed DM.  Stars and DM in debris flow are also well-correlated.  The correspondence is not as robust for younger mergers leaving behind streams, because spatial offsets between the DM and stars can lead to localized variations in their velocity components.  

We demonstrate how to recover the total DM distribution in the solar neighborhood in cases where it is dominated by a relaxed population and debris flow.  After successfully demonstrating this procedure with simulations from the \textsc{Fire} project, we apply it to the Milky Way and the recently discovered debris field in the Solar neighborhood.  This procedure pertains specifically to DM accreted from luminous satellites and therefore does not account for contributions from non-luminous satellites, which requires further study.  Additionally, the conclusions are specific to the solar circle (defined as $|r-r_{\odot}| < 2$ kpc  and $|z| \leq 1.5$~kpc with $r_\odot$ the solar radius), which is the volume studied in this work.

This paper is organized as follows. Sec.~\ref{sec:fire} introduces the \textsc{Fire} simulations and provides more details about the two host halos studied in this work.  Sec.~\ref{sec:origins} describes the breakdown of the DM and stars within the solar circle of the hosts in terms of their accretion time and progenitor characteristics.  \Sec{sec:correlation} discusses the correlation between the stars and DM for the relaxed, debris flow, and stream categories described above.  \Sec{sec:totaldarkmatter} demonstrates how to build the total DM distribution; this new strategy is applied to the Milky Way in \Sec{sec:milky_way}.  We conclude in \Sec{sec:conclusions}.  The Appendix includes additional figures that supplement the main results of the paper.

\section{FIRE-2 Simulations}
\label{sec:fire}

\subsection{The Host Halos}

We analyze two cosmological zoom-in \citep{KatzWhite1993,Onorbe2014} hydrodynamic simulations from the \emph{Latte} suite \citep{Wetzel2016} of \textsc{Fire-2} simulations \citep{2017arXiv170206148H}.
\textsc{Fire-2} simulations are run using the~\texttt{GIZMO} code\footnote{\url{http://www.tapir.caltech.edu/~phopkins/Site/GIZMO.html}} \citep{Hopkins:2014qka} with the mesh-free finite-mass (``MFM'') Lagrangian Godunov method for hydrodynamics, while gravity is solved using a version of the Tree-PM solver from \texttt{GADGET-3}~\citep{Springel:2005mi}.
We briefly review the details of these simulations that are most relevant for our study; see \cite{2017arXiv170206148H} and \cite{2018arXiv180610564S} for more details.

\textsc{Fire-2} simulations include heating from a meta-galactic background~\citep{2009ApJ...703.1416F} and cooling from local stellar sources from $T\sim10$--$10^{10}$~K.  Star formation occurs in locally self-gravitating~\citep{Hopkins:2013oba}, Jeans-unstable, self-shielding~\citep{Krumholz:2010wm} molecular gas. Stellar feedback occurs through photoionization, photoelectric heating, radiation pressure, supernovae Ia \&\ II, and stellar winds from primarily O, B and AGB stars. Inputs are taken directly from stellar evolution models using \texttt{STARBURST99~v7.0}~\citep{1999ApJS..123....3L,2014ApJS..212...14L} and assume the~\cite{Kroupa:2000iv} IMF.  The \emph{Latte} simulations that we use also include sub-grid turbulent diffusion of metals in gas \citep{2017arXiv170206148H,2017MNRAS.471..144S}, which produce more realistic metallicity distributions \citep{Escala2017}.  

We focus on the galaxies \mi~(introduced in \citealt{Wetzel2016}) and \mf~(introduced in \citealt{Garrison-Kimmel:2017zes}), which provide contrasting formation histories: the latter experiences more mergers at late cosmic times. Both \mi~and \mf~assume a $\Lambda$CDM cosmology with $\Omega_\Lambda = 0.728$, $\Omega_m = 0.272$, $\Omega_b = 0.0455$, $h = 0.702$, $\sigma_8 = 0.807$, and $n_s = 0.961$. The initial mass of baryonic particles is $7070$~M$_\odot$ (though because of stellar mass loss, the typical star particle has mass $\approx 5000$~M$_\odot$ at redshift $z = 0$); the gravitational softening length is 4~pc (Plummer equivalent) for stars and gas has adaptive softening/smoothing down to 1~pc.
DM particles in the zoom-in region have mass $3.5 \times10^4$~M$_\odot$ and softening length of 40~pc.

At redshift $z=0$, the primary host halo in \mi~has $M_\mathrm{200m} =1.2\times10^{12}$~M$_\odot$ and $R_\mathrm{200m} = 336$~kpc, defined via the radius containing 200 times the average matter density. Within this radius, the host halo contains $N_\mathrm{particle} = 5.08\times10^7$ DM, gas, and star particles. The corresponding properties for the host halo in \mf~are as follows:  $M_\mathrm{200m} =1.7\times10^{12}$~M$_\odot$, $R_\mathrm{200m} = 380$~kpc, and $N_\mathrm{particle} = 7.44\times 10^7$. Each host halo is selected to be isolated, with no equally massive halos within $5 R_\mathrm{200m}$.

The host galaxies of \mi~and \mf~are similar in many respects to the Milky Way~\citep{2018arXiv180610564S}. For example, the total stellar mass of the Galactic disk is $(5\pm1) \times 10^{10}$~M$_\odot$~\citep{2016ARA&A..54..529B}, compared to $5.5\times10^{10}$ and $6.9\times10^{10}$~M$_\odot$ in \mi~and \mf, respectively (this differs from the total mass inside $R_\mathrm{200m}$ as it excludes satellites). Additionally, these simulations provide a reasonable match to the observed morphology of Milky Way-like galaxies~\citep{2017arXiv171203966G, 2018arXiv180610564S}, disk kinematics and abundance gradients~\citep{Ma:2016fbd}, satellite dwarf galaxy stellar masses, velocity dispersions, metallicities, and star-formation histories \citep{2016ApJ...827L..23W, 2018arXiv180604143G, Escala2017}, and properties of the thick disk and stellar halo \citep{2017arXiv171205808S, 2017arXiv170405463B}.

We identify DM (sub)halos using the \textsc{Rockstar} phase-space finder\footnote{\url{https://bitbucket.org/pbehroozi/rockstar-galaxies}}~\citep{Behroozi:2011ju}, and we generate merger trees using \textsc{ConsistentTrees} (REF) across 600 snapshots from redshifts $z=0$--99.
We ran the halo finder on only the DM particles, and we assigned stars to each halo in post-processing (see below).

\subsection{Tracking Dark Matter and Stars}

To understand the origin of stars and DM near the solar circle, we track the location of DM/star particles over all snapshots.  To start, we identify all the DM particles in the solar circle of the host ($|r-r_{\odot}| < 2$ kpc  and $|z| \leq 1.5$ kpc) at the present day. We then follow the location of every particle at each previous snapshot, checking if it falls within the virial radius $R_\mathrm{200m}$ of a (sub)halo and if its velocity lies within 3$\sigma$ of the (sub)halo's internal velocity (\emph{i.e.}, the maximum between its maximum circular velocity and its velocity dispersion). If these conditions are met, we mark the (sub)halo as the particle's host, further requiring that the DM is associated with the same (sub)halo for 6 out of the last 9 snapshots to avoid contamination by flybys that happen to fall within the velocity dispersion.
We mark $\zacc$ as the last redshift at which the particle was bound to the (sub)halo; the particle is bound to the primary host halo in the following snapshot. These requirements lead to an unassociated DM fraction of 69\%~(74\%) for \mi~(\mf).

\begin{table*}[t]
\centering
\footnotesize
\renewcommand{\arraystretch}{1.5}
\begin{tabular}{C{4.3 cm} |  C{1.4cm} C{1.4cm} C{1.4cm} |C{1.4cm} C{1.4cm} C{1.4cm}}
\Xhline{3\arrayrulewidth}
&\multicolumn{3}{c|}{\textbf{\texttt{FIRE m12i} Host Halo}} &  \multicolumn{3}{c}{\textbf{\texttt{FIRE m12f} Host Halo}} \\
 &  I & II & III  &  I & II & III    \\ 
 \hline
 $M_\mathrm{peak}$~[M$_\odot$] &  $6.5\times10^{10}$ & $3.6\times10^{10}$ & $3.8 \times 10^{10}$ &  $1.5  \times 10^{11} $ & $8.1 \times10^{10}$ & $3.2\times10^{10}$  \\
  $\langle \FeH \rangle$ &  $-1.47$ & $-1.82$ & $-1.85$ &   $-0.90$ & $-1.14$ & $-1.83$  \\
 $M_\mathrm{peak}/M_\mathrm{*, total}$  & 122 & 101 & 228  &  82 & 66  & 162  \\
Stellar Mass Fraction &  34\% & 24\% & 22\% & 47\% & 34\% & 6.0\%  \\
Dark Matter Mass Fraction &   24\% & 32\% & 14\% &  23\% & 33\% & 8.0\%  \\
Stellar Accretion Redshift $(z_\mathrm{acc})$ &  1.07--1.70 & 2.06--2.27 & 2.90--3.30 &  0.17--0.39 & 0.73--0.94 & 3.70--3.80  \\
 $M_\mathrm{DM}/M_\mathrm{*}$ at Solar Circle &  19 & 35 & 18 &   6 & 12 & 17 \\
    \Xhline{3\arrayrulewidth}
  \end{tabular}
  \caption{Properties of the top three mergers (labeled as I--III) in \mi~and~\mf, ranked by the fraction of accreted stellar mass each contributes to the solar circle.  For each galaxy, we list the peak mass of its dark matter halo ($M_\mathrm{peak}$), average stellar metallicity ($\langle \FeH \rangle$), and peak halo-to-stellar mass ratio ($M_\mathrm{peak}/ M_\mathrm{*, total}$).  We also provide the stellar and dark matter mass fractions contributed by each  satellite galaxy within the solar circle.  Note that all fractions are taken with respect to the total accreted material from subhalos with $M_\mathrm{peak} > 10^9$~M$_\odot$ in the simulation.  The range of accretion redshifts ($\zacc$) for the stars that are stripped from each satellite is also listed.  The final row corresponds to the ratio of dark matter mass to stellar mass contributed by each satellite within the solar circle ($|z| \leq 1.5$~kpc and $r_{\odot} \pm 2$~kpc, where $r_\odot = 8$~kpc).}
  \label{tab:progenitors}
  \end{table*}

The procedure to associate stars to each subhalo is similar. A star particle must lie within a subhalo's virial radius and have a velocity that falls within 2.5$\sigma$ of the subhalo's stellar velocity dispersion (computing membership and velocity dispersion iteratively until convergence). We include as `galaxies' only subhalos that contain at least 10 stars.
We also require that a star particle is part of the same subhalo for at least 3 out of the last 4 snapshots.\footnote{Because stars are born from gas, requiring them to be associated for 6 out of 9 snapshots like the DM could bias us towards an older stellar population.}
We quote the stellar mass of a given subhalo at the particle's $\zacc$.

In this manner, we identify the subhalo progenitor of each DM/star particle observed today in the solar circle of the primary host galaxy.  We also store information on the progenitor subhalo, such as its total DM and stellar mass. Because of tidal stripping, the total mass of a subhalo at $\zacc$ is typically smaller than its initial mass before falling into the primary host.
Thus we also use the subhalo peak mass, $M_\mathrm{peak}$, computed from the merger trees.

There are two important resolution effects that affect our ability to track all the DM and star particles in the solar circle. First, there is a minimum mass for luminous subhalos in the simulation set by the mass of each star particle ($\sim 5000$~M$_{\odot}$ at redshift $z = 0$). Because we only track galaxies with at least 10 star particles, this leads to an effective lower limit on the total stellar mass of a satellite to be $\sim 10^5$~M$_\odot$, which corresponds to a halo mass of $\sim 5 \times 10^{8}$~M$_\odot$.  Thus, we conservatively label the subset of subhalos with $M_\mathrm{peak} \gtrsim 10^9$~M$_{\odot}$ to be luminous in this work.

Second, there is a minimum (sub)halo mass of $\sim 10^6$~M$_\odot$ because of the DM mass resolution. When tracking the origin of a DM particle, we may find that it is not associated with a specific progenitor. This may either be because its (sub)halo is not resolved or because the DM was never associated with a (sub)halo and was accreted smoothly.
We cannot distinguish between these two possibilities.

Throughout the paper, we will separate the DM into two components. The first is the component that originates from luminous subhalos with $M_\mathrm{peak} > 10^9$~M$_\odot$. The second is the component that originates from either a subhalo whose galaxy was not adequately resolved, a dark subhalo, an unresolved subhalo, or smooth accretion.  We will refer to this component as `Dark/Unresolved.'

\section{Accretion History at the \\Solar Circle}
\label{sec:origins}

Because the primary focus of this work is the local DM velocity distribution, we restrict the study of \mi~and \mf~to the volume within distances $|z| \leq 1.5$~kpc of the midplane and galactocentric radii $r_{\odot} \pm 2$~kpc, where $r_\odot = 8$~kpc. This is justified because the scale radii of the simulated disks are comparable to those of the Milky Way \citep{2017arXiv171205808S}.  We refer to this volume as the `solar circle.'  There are a total of $\sim 1.70\times 10^5$ ($2.19\times10^5$) DM and $\sim 9.78\times10^5$ ($1.48\times10^6$) star particles within this region of \mi~(\mf). 

The total fraction of accreted stars at redshift $z=0$ constitutes only $1.5\%$ ($2.2\%$) of all stars in the solar circle of \mi~(\mf).\footnote{Note that when we refer to `accreted stars,' we do not include stars that formed from gas that accreted onto the host early on.}  The vast majority of the stars are born \emph{in-situ}---that is, they are born within the host galaxy~\citep{2009ApJ...702.1058Z,2011MNRAS.416.2802F,2012MNRAS.420.2245M,Pillepich:2014jfa,2015MNRAS.454.3185C,2017arXiv170405463B}.  However, the fraction of accreted stars increases towards lower metallicities.  The probability of a star being accreted with a metallicity $\FeH < -2$ is $66\%$ ($89\%$) for \mi~(\mf).  This increases to $95\%$~($99\%$) for \mi~(\mf)~when requiring $\FeH < -3$.

Table~\ref{tab:progenitors} lists the top three satellite galaxies that contribute the greatest fraction of \emph{accreted} stellar mass at the solar circle of \texttt{m12i}.  We see that 34\%  of these stars were accreted between redshifts of $\zacc = 1.07$--1.70 from a $6.5 \times 10^{10}$~M$_\odot$ satellite. 
The next 24\% of stars were accreted at $\zacc=2.06$--2.27 from a $3.6\times 10^{10}$~M$_\odot$ satellite.  In contrast, the majority of the local stellar halo in \mf~formed at lower redshifts.  
For example, nearly half of the stellar mass at the solar circle today was accreted between $\zacc=0.17$--0.39.  

Because the dominant mergers in \mf~are typically younger relative to those of \mi, they are more luminous and have a smaller ratio of peak mass to stellar mass with $M_\mathrm{peak}/M_\mathrm{*, total}= 66$--162 compared to 101--228  for \mi.  This also leads to a more metal-rich population of accreted stars for \mf~relative to \mi, with mean metallicities of the dominant mergers closer to $\langle \FeH \rangle_\mf \sim -1.3$ compared to $\langle \FeH \rangle_\mi \sim -1.7$. 

Mergers I--III contribute nearly all of the local \textit{accreted} stellar mass in \mi~and~\mf, and a comparable fraction of the accreted DM.  `Accreted DM' refers to the DM that originates from subhalos with $M_\mathrm{peak} > 10^9$~M$_\odot$, and excludes the `Dark/Unresolved' component.  In \mi, for example, $80\%$ of the accreted stellar mass comes from Mergers I--III, whereas $70\%$ of the accreted DM does. In \mf, the top three mergers contribute $87\%$ of the accreted stars and $64\%$ of the accreted DM.  

\begin{figure*}[tb] 
   \centering
		\includegraphics[width=0.45\textwidth]{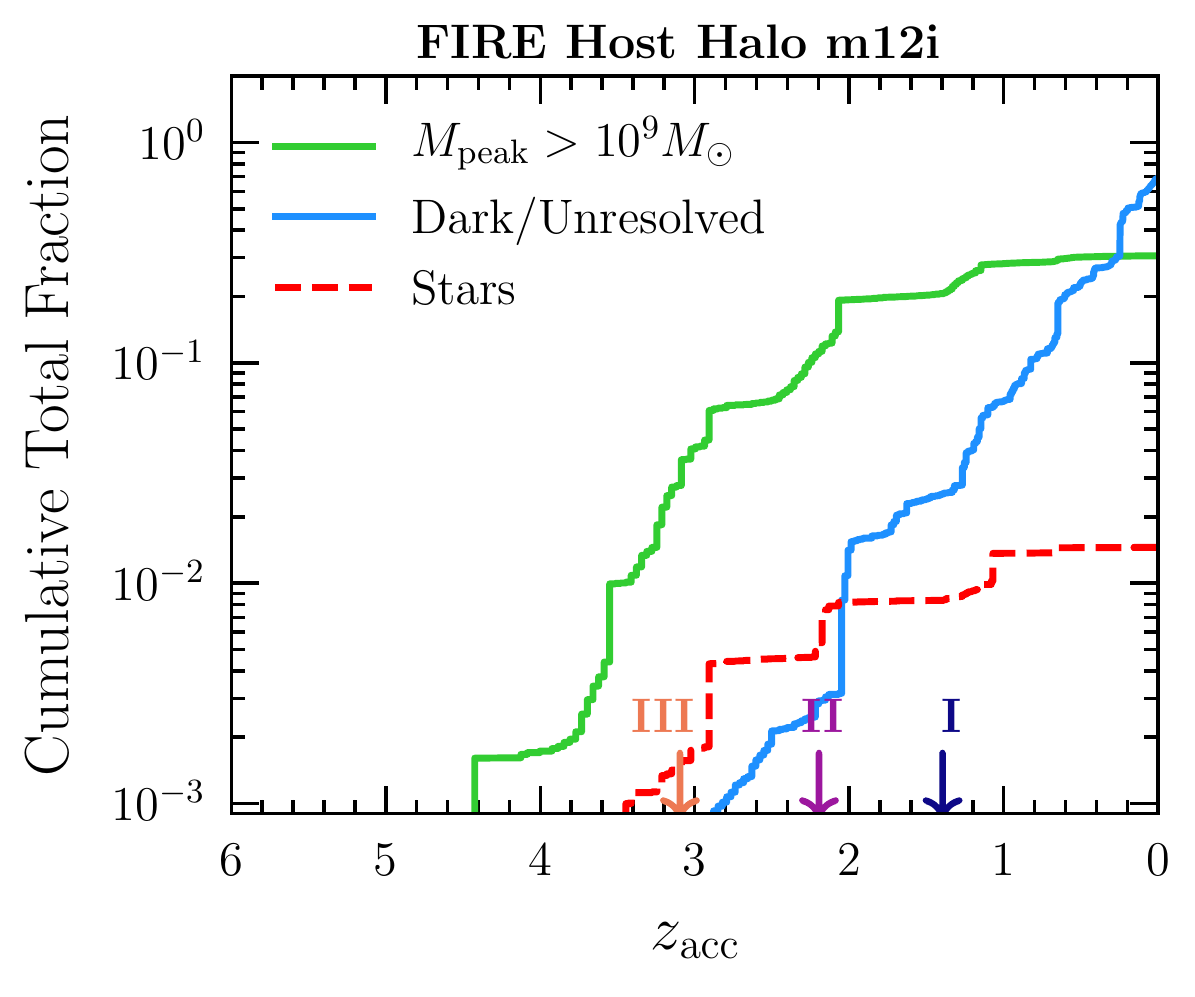} 
		\qquad
		\includegraphics[width=0.45\textwidth]{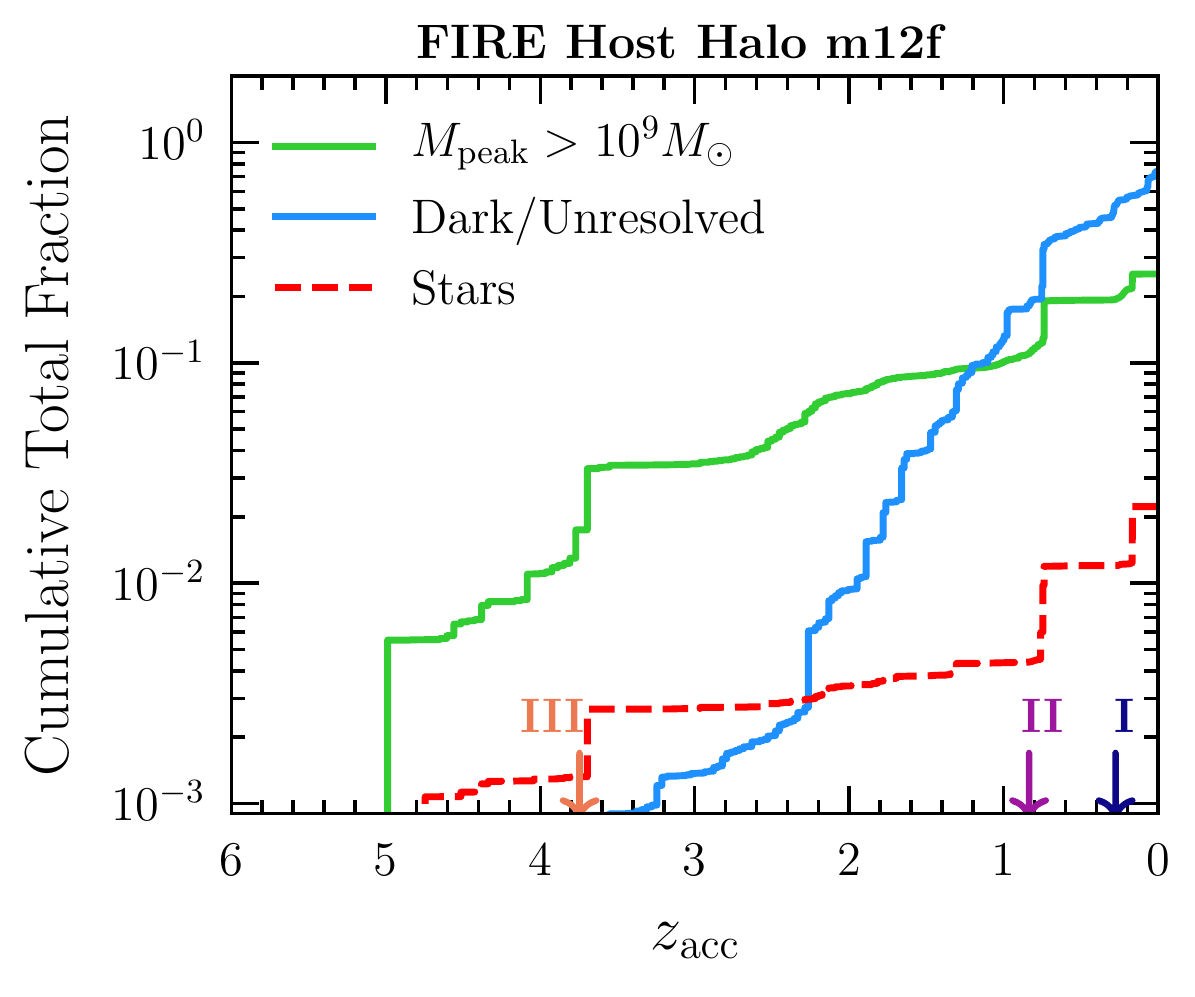} 
   \caption{The cumulative fraction of dark matter and stars at the solar circle of simulated host \texttt{m12i} (left) and \texttt{m12f} (right).  The dark matter is divided into two separate contributions. The first (green solid) is from luminous satellite galaxies with peak halo mass $M_\text{peak} > 10^9$~M$_\odot$.  The second (aqua solid) is dark matter that originates from either a subhalo whose galaxy was not adequately resolved, a truly dark subhalo, an unresolved subhalo, or smooth accretion; due to the finite mass resolution of the DM and star particles in the simulation, it is not possible to further distinguish its origin.  The dashed red line corresponds to the cumulative fraction of accreted stars.  The cumulative fraction is defined with respect to the total number of particles of each kind found in the solar circle at redshift $z=0$.  The deficit below unity at $\zacc = 0$ for the stellar distribution corresponds to its \emph{in-situ} fraction.  
}
   \label{fig:origins}
\end{figure*}

\Fig{fig:origins} shows the cumulative fraction of DM as a function of accretion redshift for \mi~(left) and \mf~(right).  We separately show the total DM that was accreted from galaxies with $M_\mathrm{peak} > 10^9$~M$_\odot$ in green and the `Dark/Unresolved' component in aqua.  As discussed in Sec.~\ref{sec:fire}, $M_\text{peak} \sim 10^9$~M$_\odot$ is roughly the lower limit for luminous satellites in the simulation given the resolved star particle mass. Luminous satellites in the simulation with halo masses above this limit offer an opportunity to compare the final positions of accreted stars and DM.  

\Fig{fig:origins} shows the cumulative fraction of the stars accreted from these satellites in dashed red.  The distinct steps in the cumulative stellar fraction occur at the average $\zacc$ for stars stripped from Mergers~I--III (indicated by the arrows in the figure).  Similar steps are observed in the cumulative DM fraction at roughly the same redshifts.  This explicitly demonstrates that the mergers dragged in significant amounts of both DM and stars to the solar circle at approximately the same times.  

The fact that the jumps in the DM cumulative fraction closely align with those in the stars suggests that it is the most bound DM of each satellite that contributes at the solar circle.  In general, we expect that tides start to remove DM from a satellite earlier than its stars because the halo is more extended.   By the time the satellite's orbit sinks down to the inner parts of the galaxy, however, most of its DM halo has been stripped off, leaving behind only the most bound portion.  This is confirmed by looking at the overall ratio of DM mass to stellar mass contributed by Mergers~I--III to the solar circle (bottom row of \Tab{tab:progenitors}).  Importantly, these ratios are roughly an order of magnitude below  $M_\mathrm{peak}/M_\mathrm{*, total}$, suggesting that a large fraction of the halo's DM has already been removed by the time it has sunk to the inner parts of the host galaxy.  This is a crucial observation, as it suggests why the DM and stars from these mergers should  share similar kinematics near the solar position.  By the time a massive satellite passes near the sun, its outer halo has mostly been stripped away, and the DM being removed is concentrated near the central parts of the satellite, similar to the stars.  In this respect, the sun's location at the inner galaxy is fortuitous for reconstructing the DM velocities from stellar orbits.

\section{Correlations Between Accreted Stars and Dark Matter}
\label{sec:correlation}

The phase-space distribution of the DM and stars within the solar circle is intimately linked with the galaxy's accretion history.  DM and stars that accreted onto the host at early epochs ($\zacc \gtrsim 3$) are fully relaxed.  More recent accretion events, however, continue to build up the local mass profile.  If this debris is not fully phase mixed, it can be identified as substructure in either position or velocity space. 
 
 \begin{figure*}[tb] 
   \centering
	\includegraphics[width=0.95\textwidth]{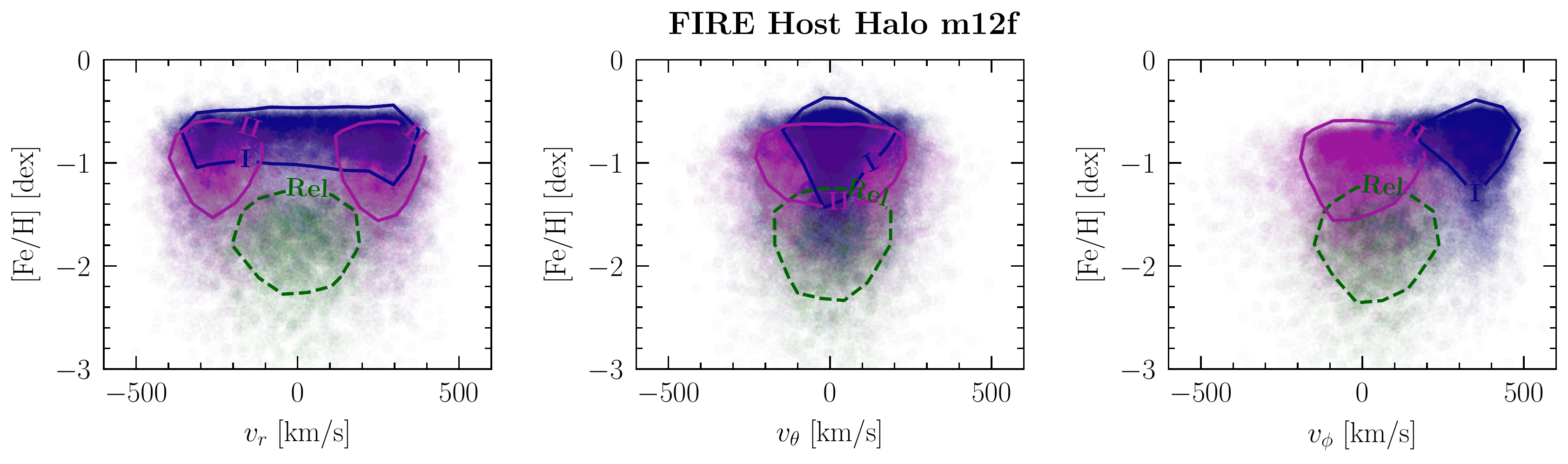} 
	\includegraphics[width=0.95\textwidth]{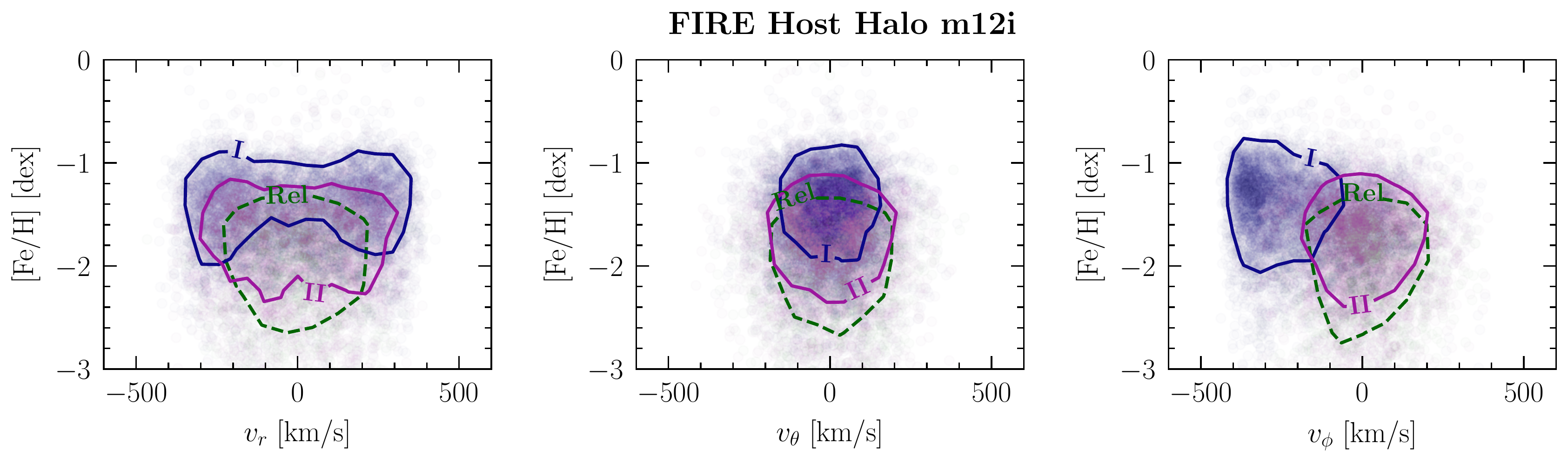} 
   \caption{The 66\% containment region in metallicity-velocity space for stars within the solar circle of \mf~(top) and \mi~(bottom) that are stripped from Mergers~I and~II (blue and pink solid, respectively).  We also show the corresponding distributions for the relaxed component (green dashed), defined as the subset of stars accreted before redshift $\zacc > 3$.  Note that Merger~III is included in this population.   Velocities are in spherical Galactocentric coordinates, with $\phi$ the azimuthal direction aligned with the disk rotation. }
   \label{fig:all_mergers}
\end{figure*}

 \Fig{fig:all_mergers} demonstrates how the stars in both the relaxed and substructure populations cluster in metallicity-velocity space.   In general, elemental abundances provide an important handle when linking stellar debris to a progenitor galaxy~\citep{Johnston:1995vd, Johnston:1996sb, Helmi:1999ks, Bullock:2000qf, Bullock:2005pi, Purcell:2007tr, DeLucia:2008gk}; we focus on the iron abundance $\FeH$ here.  
\Fig{fig:all_mergers} shows the distributions of $\FeH$ against $v_r, v_\theta, v_\phi$ for stellar debris of \mf~(top) and \mi~(bottom).  Note that we use spherical Galactocentric velocities throughout, with $\phi$ oriented with the disk rotation.  The relaxed stellar component is shown in green, while the stellar populations associated with Mergers I~and~II are shown in blue and pink, respectively.  Merger~III is included in the relaxed population.  Clearly, a wide variety of kinematic features are possible.  While the relaxed stellar population appears to be nearly isotropic, the more recent mergers exhibit distinctive kinematic features.  Taken together, the chemical abundance and kinematics of stellar populations can play an important role in identifying their origin.  
 
In this section, we explore in detail the phase-space evolution of DM and stars from mergers in \mi~and \mf.  We systematically study the contributions to the solar circle, from the oldest to the youngest accreted material.  In this way, we will see how the velocity distribution of the accreted stars is built up as a function of time, and how well it traces the DM as the two evolve and grow together.  Host halo \mf~provides a contrasting example to \mi, because its merger history is more active up until redshift $z\sim0.3$.

The results of this section pertain specifically to DM that is sourced by luminous satellites ($M_\text{peak} \gtrsim 10^9$~M$_\odot$).  The kinematic distributions of DM from the `Dark/Unresolved' component is discussed in \Sec{sec:untracked}.

\subsection{The Relaxed Component}
\label{sec:virialized}

\begin{figure*}[tb] 
   \centering
   	\includegraphics[width=0.95\textwidth]{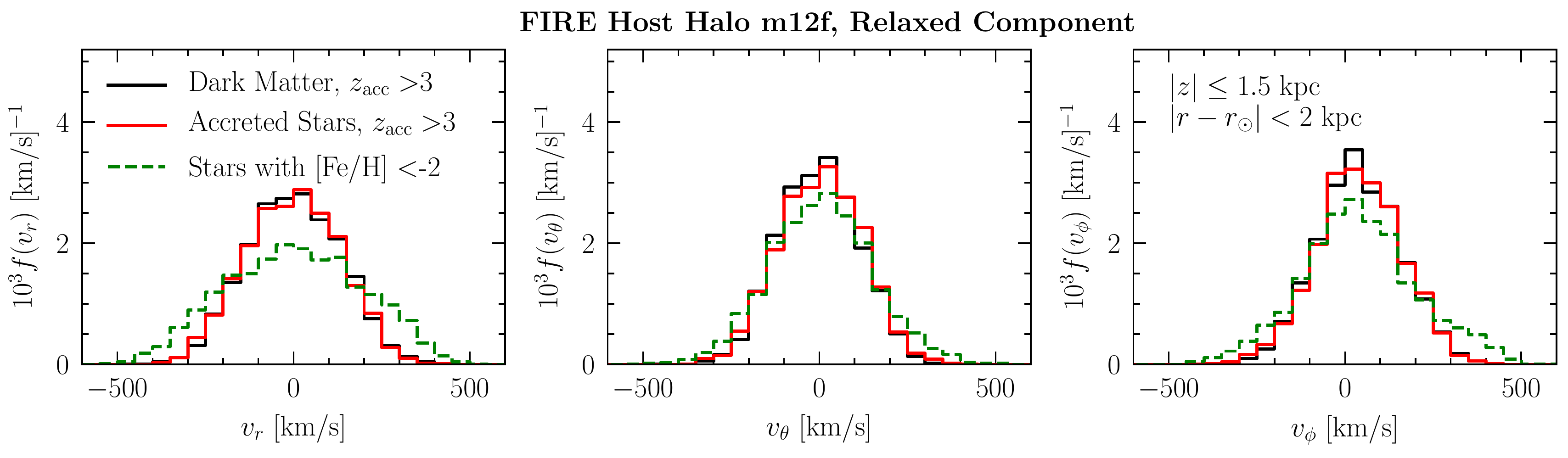} 
	\includegraphics[width=0.95\textwidth]{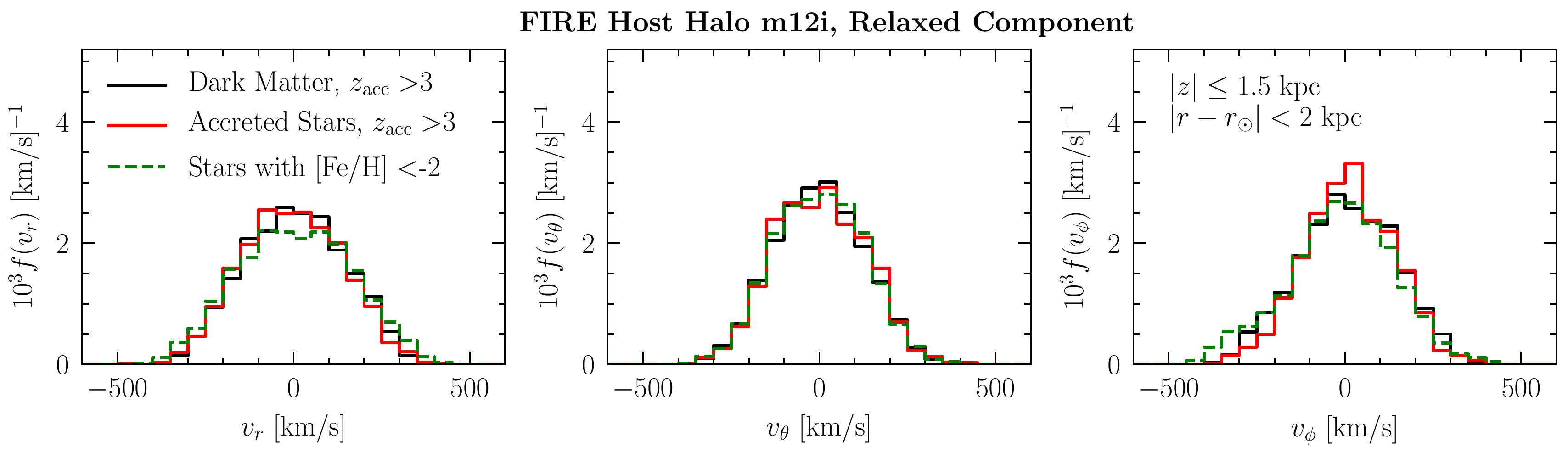} 
   \caption{Present-day velocity distributions for the stars (red solid) and dark matter (black solid) accreted before redshift $z_{\rm{acc}} > 3$ in \mf~(top) and \mi~(bottom).  We also show the corresponding distributions for all stars (not just the  accreted subset) with $\FeH <-2$ (green dashed).  The discrepancy between the low-metallicity stellar sample and the relaxed dark matter distribution in the radial distribution of \mf~is due to contamination by Merger~II below $\FeH \lesssim -2$.  Applying more sophisticated clustering algorithms to the stellar data could help reduce such contamination.  \Fig{fig:virialized_m3} of the Appendix shows the corresponding distributions for $\FeH \lesssim -3$.  }
   \label{fig:virialized}
\end{figure*}

Violent relaxation plays an important role in mixing stars and DM that accreted from a galaxy's oldest mergers.  Non-adiabatic transformations of the potential change the energies of the stars and DM, causing their orbits to fill  the available phase space.  These effects are particularly important in the period when the proto-Milky Way is forming.  This process is distinct from changes to the course-grained phase-space distribution that arise as a system evolves in time following Liouville's theorem.  In this process, both the original phase-space volume and energy are conserved as time evolves.  This phase-mixing process drives the evolution of streams and debris flow, as described in Sec.~\ref{sec:debris}.      

We begin by focusing on the present-day distribution of stars and DM in \mi~and~\mf~that were accreted from the earliest mergers ($\zacc >3$).  There are 21 significant mergers that contribute to this population in \mi, and 34 for \mf.  Note that the relaxed population in both hosts includes Merger~III.  The average metallicity of the stars from these mergers is  $\langle \rm{[Fe/H]} \rangle_{\mi}  = -2.04$ (0.52 dex spread) for \mi, and $\langle \rm{[Fe/H]} \rangle_{\mf}  = -1.89$ (0.48 dex spread) for~\mf.   

The velocity distributions of the relaxed stellar component in \mi~is indicated by the red lines in the bottom panel of~\Fig{fig:virialized}.  The distributions are approximately isotropic, with dispersions of $\{ \sigma_r, \sigma_{\theta}, \sigma_{\phi} \} =  \{139, 127, 125 \}$~km/s.  Notably, the stellar and DM distributions, which are indicated in black, trace each other closely. The discrepancies between the two are small, ranging from $0.5$--$17$\% in any given bin, but closer to $\sim 50\%$ along the tails.  As the top panel of \Fig{fig:virialized} shows, these results are similar for \mf.

\begin{figure*}[tb] 
   \centering
	\includegraphics[width=0.95\textwidth]{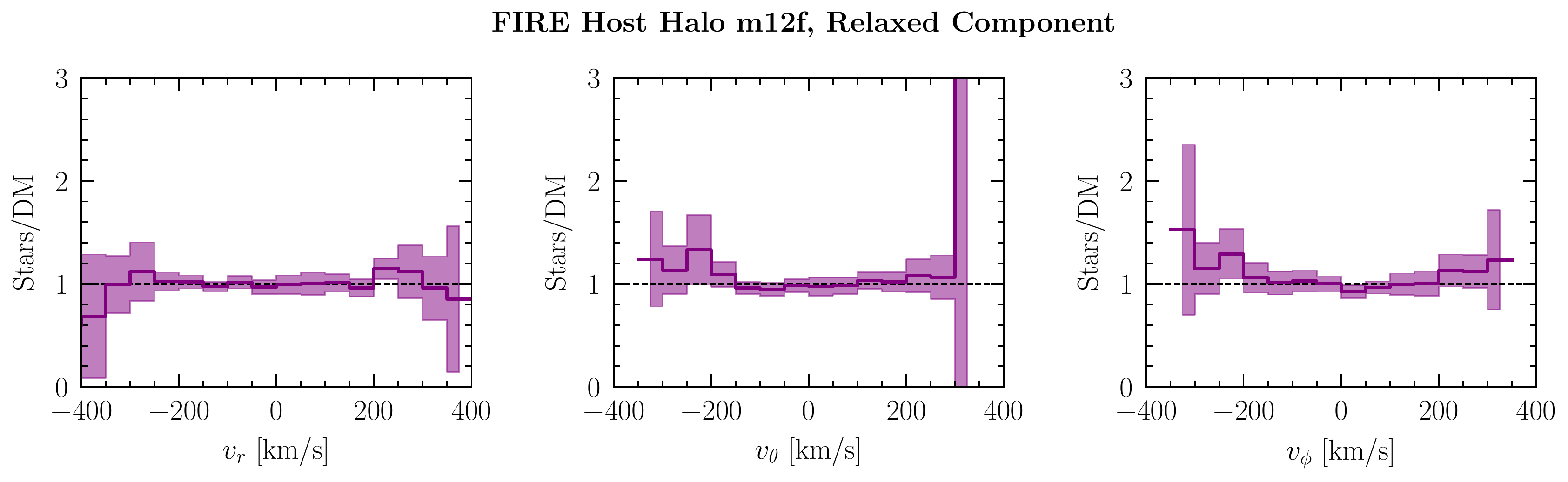} 
	\includegraphics[width=0.95\textwidth]{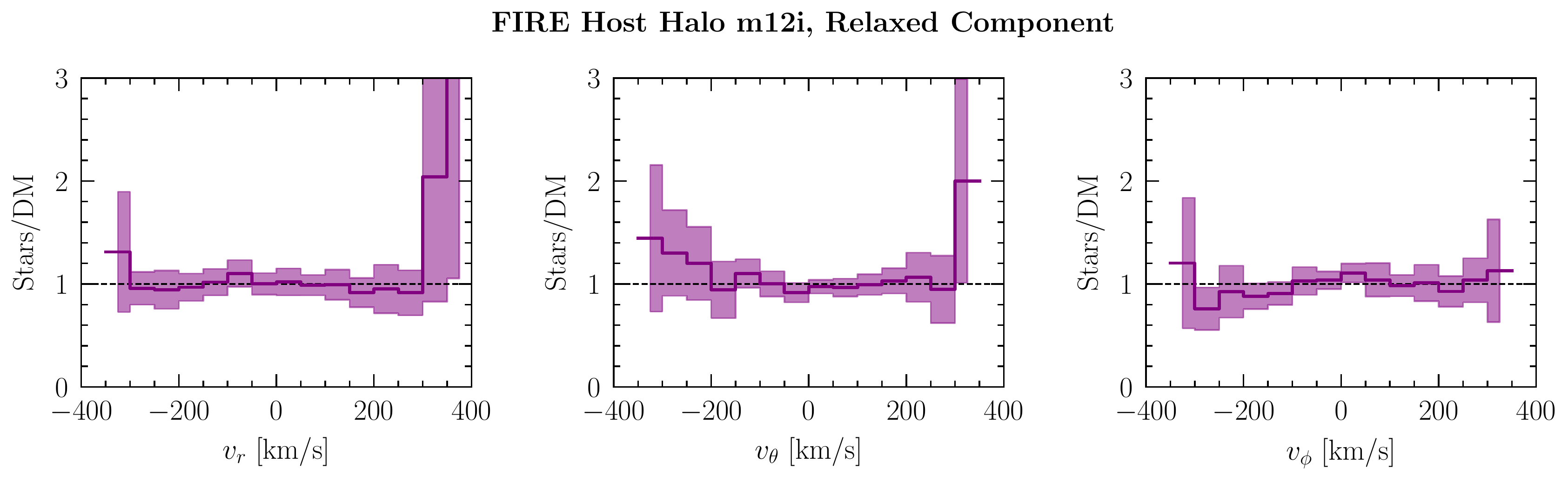} 
   \caption{The ratio of the stellar to dark matter (DM) velocity distributions for the relaxed population of \mf~(top) and \mi~(bottom).  Results are shown separately for the separate Galactocentric velocity components.  The distributions are sampled in 10 locations throughout the solar circle, within spheres of radius 4~kpc centered at a Galactic distance of $r_\odot = 8$~kpc.  The mean ratio over these regions is indicated by the solid purple line and the colored band indicates the 1$\sigma$ spread.  }
   \label{fig:ratio_merger_vir}
\end{figure*}

Using the \texttt{Eris} simulation, \cite{Herzog-Arbeitman:2017fte} demonstrated that metal-poor stars act as kinematic tracers for the relaxed DM component.\footnote{Note that what we refer to as `relaxed' here is referred to as `virialized' in \cite{Herzog-Arbeitman:2017fte}. The \texttt{Eris} study did not break down the DM into components from older versus more recent mergers.  The fact that a good correspondence was already observed with metal-poor stars in this case suggests that the shape of the local DM distribution in that host was not significantly affected by substructure, dark subhalos, and/or smooth accretion. }  To test whether the same results are reproduced with \texttt{Fire},  we compare the relaxed distributions to those of all stars (not just the accreted subset) with a metallicity cut of $\FeH<-2$ (green dashed).  For \mi, the metal-poor stars trace the relaxed component of DM and stars almost exactly.  The correspondence for \mf~is also very good, especially for $v_\theta$ and $v_\phi$.  For the radial distribution, the distribution of metal-poor stars is clearly more extended.  This arises from contamination of the high-radial velocity lobes of Merger~II, which extend below $\FeH < -2$ (see \Fig{fig:all_mergers}). Tightening the metallicity requirement to $\FeH < -3$ brings the metal-poor distributions even more in-line with the relaxed distributions---see \Fig{fig:virialized_m3} of the Appendix.  

In practice, it is possible to reduce the contamination of more recent mergers, such as Merger~II of \mf, to the reconstructed distributions of the relaxed population.  A more sophisticated clustering algorithm, such as that performed in~\cite{necib2018}, can group stars based both on their metallicities and velocities.  Applied to the local stellar halo of \mf, for example, such a procedure could potentially distinguish the stars with $\FeH < -2$ that are kinematically more similar to Merger~II versus the relaxed population.

\Fig{fig:ratio_merger_vir} shows how the ratio of the relaxed stellar to DM velocity distributions varies across the solar circle.  We sample the stars and DM in spheres of radius 4~kpc that are centered at a Galactic distance of $r_\odot = 8$~kpc.  The solid purple line in \Fig{fig:ratio_merger_vir} denotes the mean value over ten sampled locations, and the band indicates the $1\sigma$ spread.  For each velocity component, the mean is consistent with unity over the majority of the velocity range, with small overall spread between regions.  Discrepancies are typically $\lesssim 10\%$, but increase to $\sim 50\%$ in the largest velocity bins, where the statistics are limited.  These results underline the fact that the DM-stellar correlation observed for the relaxed population is consistent in localized regions throughout the solar circle.

\subsection{Substructure Component}
\label{sec:debris}

\begin{figure*}[tb] 
   \centering
	\includegraphics[width=0.95\textwidth]{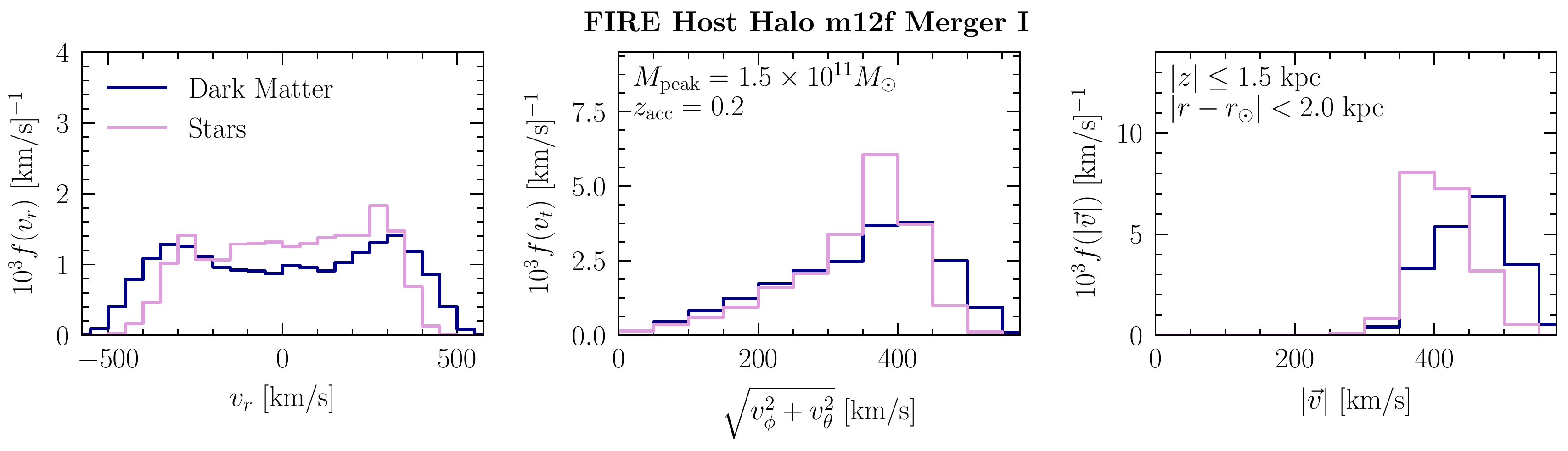} 
	\qquad
	\includegraphics[width=0.95\textwidth]{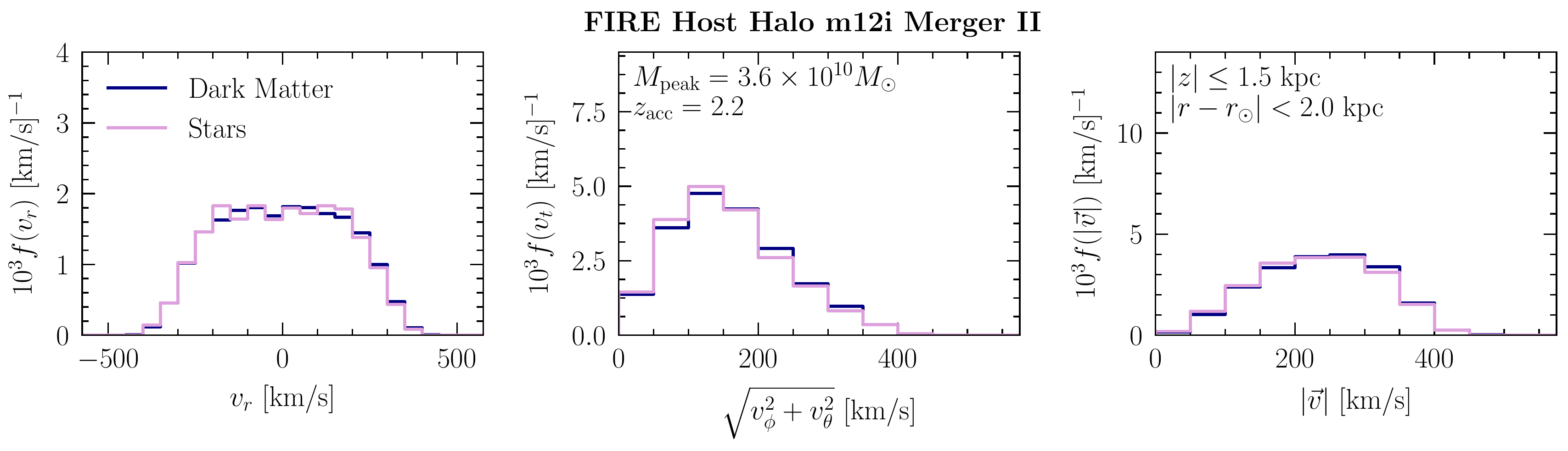} 
   \caption{Present-day velocity distributions for the debris of Merger~I of \texttt{m12f} (top) and Merger~II of \texttt{m12i} (bottom) that falls within the solar circle.  The radial (left), tangential (middle), and speed (right) distributions are shown for the stars (purple solid) and dark matter (blue solid).  The details of the mergers are provided in Table~\ref{tab:progenitors}; the corresponding distributions for the other mergers listed in the table are provided in \Fig{fig:other_mergers} and \Fig{fig:other_mergers_mf} of the Appendix.  As discussed in the text, Merger~I of \mf~is an example of a stream, while Merger~II of \mi~is an example of debris flow.}
   \label{fig:streamdebris}
\end{figure*}

\begin{figure*}[tb] 
   \centering
	\includegraphics[width=0.75\textwidth]{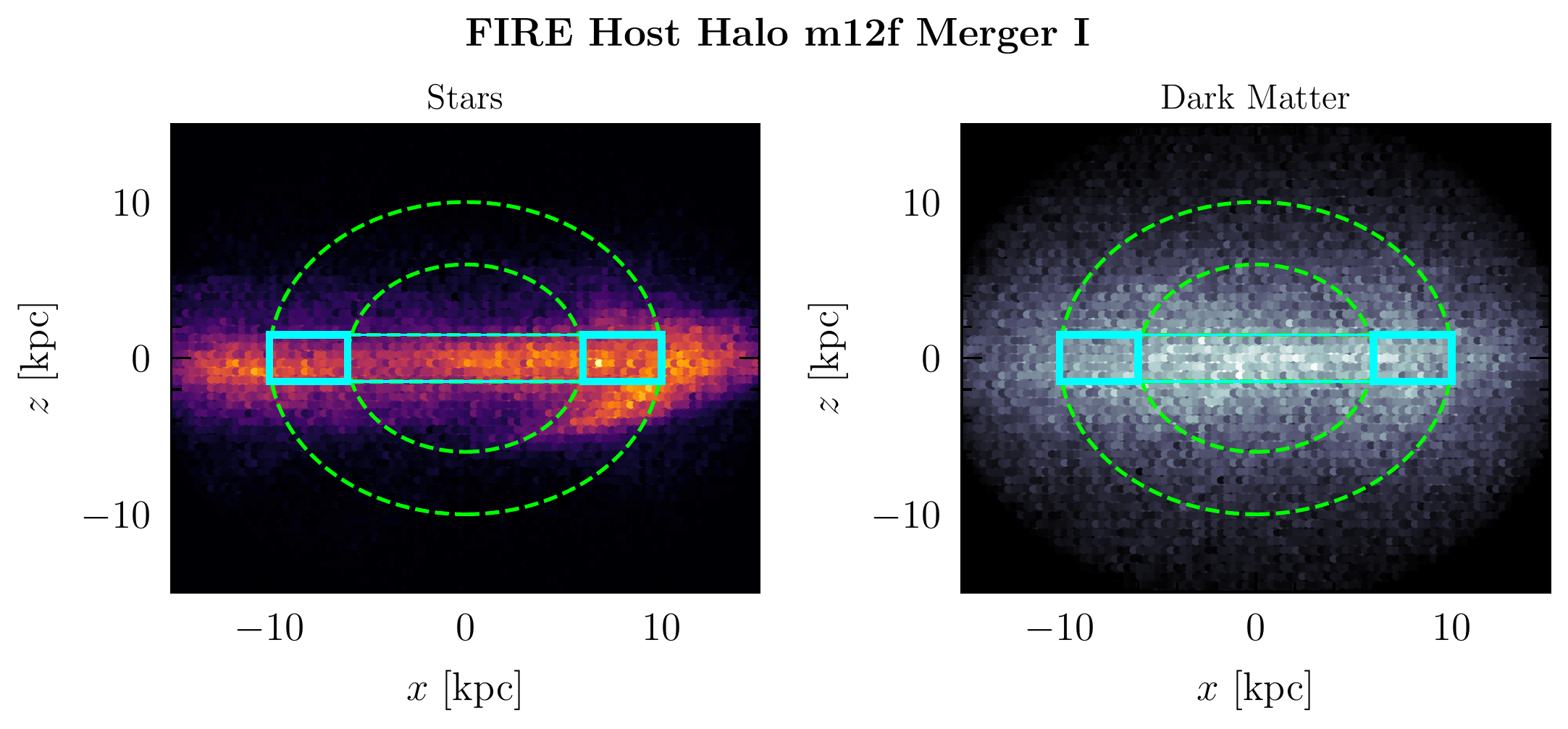} 
	\includegraphics[width=0.75\textwidth]{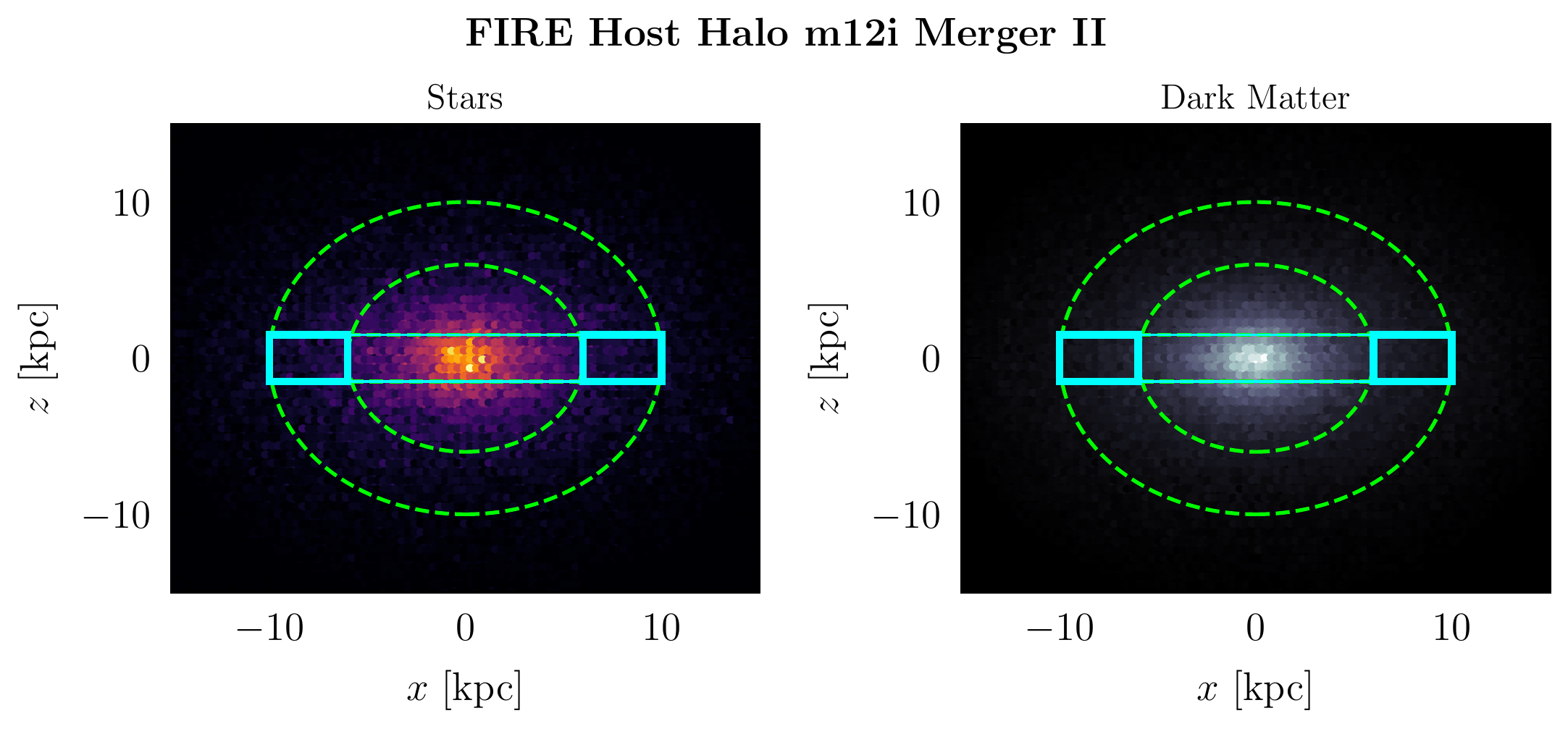} 
   \caption{Present-day spatial density distribution in the $x-z$ plane for the stars (left) and dark matter (right) from Merger~I of \mf~(top) and Merger~II of ~\mi~(bottom).  In each panel, the dashed circle corresponds to the region $|r-r_{\odot}| < 2$~kpc while the dashed green rectangle corresponds to $|z|< 1.5$~kpc.  The intersection of these two regions, denoted by the solid blue rectangle, is the solar circle. }
   \label{fig:position_merger}
\end{figure*}

After a host galaxy's last major merger, its potential changes adiabatically as DM and stars continue to be accreted through relatively smaller mergers.  
The material that is stripped is initially confined to a small region in phase space, but it evolves with time to eventually become fully mixed.  The observable features of this debris depend on the elapsed time since the merger.  For example, when the time $t$ since accretion is on the order of the dynamical time of the system ($t\sim t_\mathrm{dyn}$),  the remains of a disrupted satellite are not well-mixed either spatially or kinematically and manifest as a stream, a structure that is dynamically cold and typically coherent in speed.   Stellar streams have been observed throughout the Milky Way halo---see~\cite{2016ASSL..420...87G} and references therein---with the most studied example coming from Sagittarius~\citep{Ivezic:2000ua, Yanny:2000ty}.  DM streams have been studied in numerous $N$-body simulations~\citep{Zemp:2008gw, Vogelsberger:2008qb, Diemand:2008in,Kuhlen:2009vh,Maciejewski:2010gz,2011MNRAS.413.1419V, Elahi:2011dy}.

The most significant merger within the solar circle of \mf~leaves behind a stream.  The top row of \Fig{fig:streamdebris} shows the radial and tangential velocity distributions, as well as the speed distribution, for the DM and stars from this merger.  The stellar distribution (purple) is broad in the radial direction, while its tangential distribution is peaked at $\sim 400$~km/s.  The stars are reasonably coherent in speed, as demonstrated in the right-most panel.  The corresponding DM distributions are shown in blue.  While the DM and stellar kinematics share similar features, they do not trace each other exactly.  For example, the discrepancies between the stellar and DM speed distributions are within $3-80\%$, but reach a factor of $\sim 2$--4 at the tails.

\begin{figure*}[tb] 
   \centering
	\includegraphics[width=0.95\textwidth]{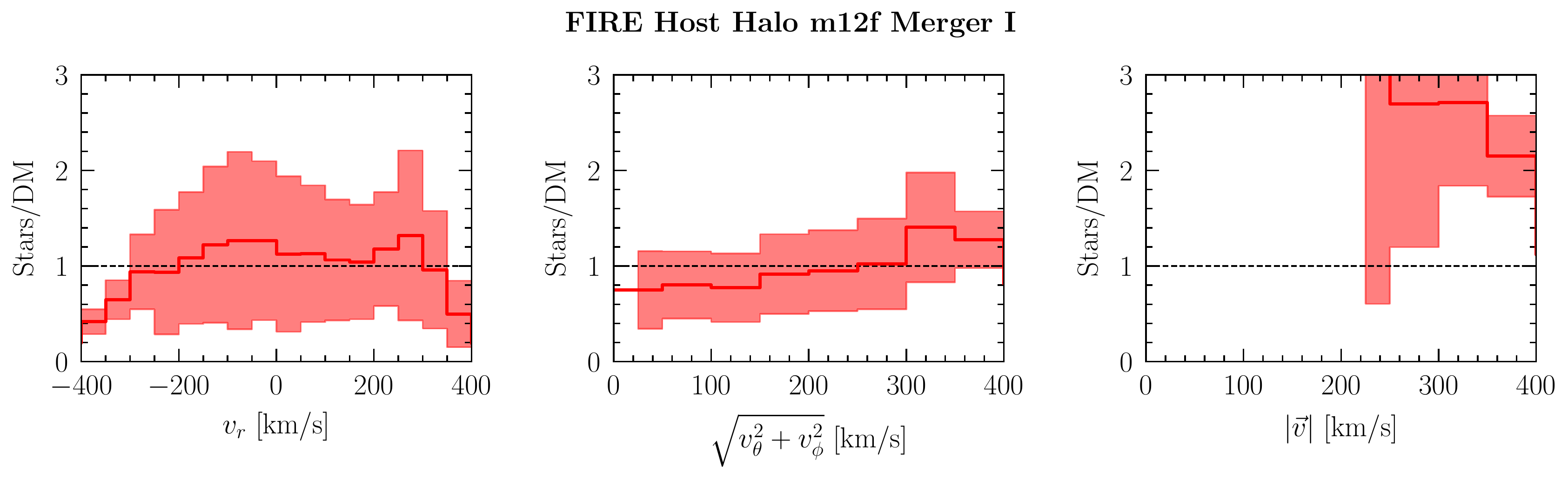} 
	\includegraphics[width=0.95\textwidth]{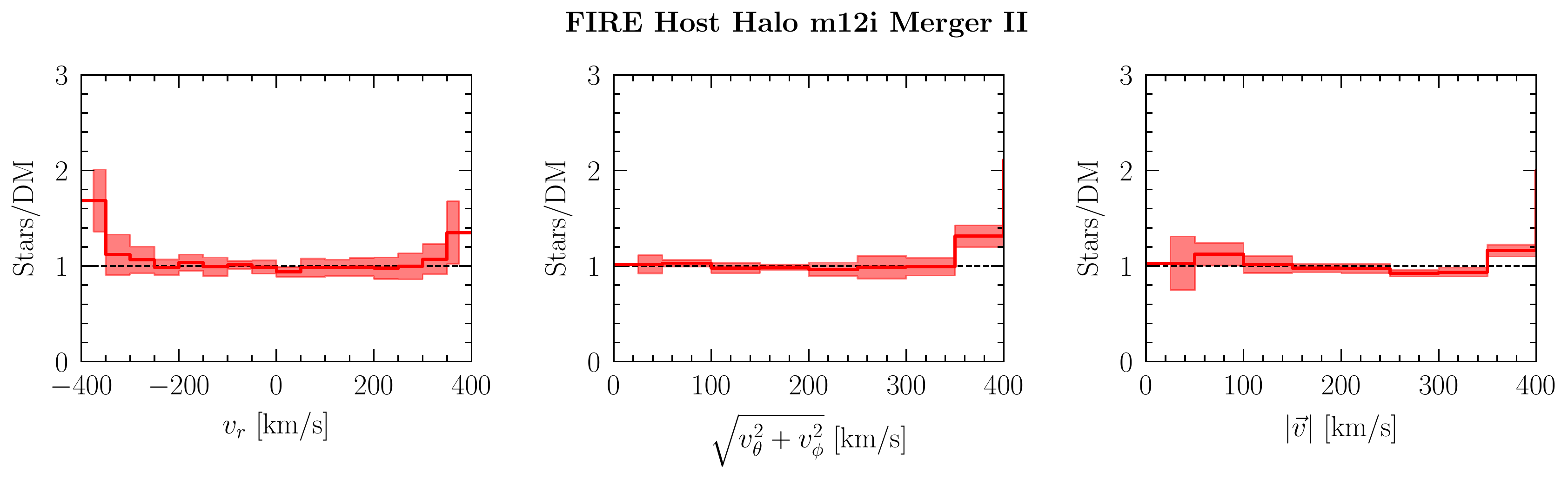} 
   \caption{Same as \Fig{fig:ratio_merger_vir}, except for the radial, tangential and speed distributions of Merger~I of \mf~(top) and Merger~II of \mi~(bottom).  }
   \label{fig:ratio_merger}
\end{figure*}

The top panel of Fig.~\ref{fig:position_merger} shows the spatial distribution of the stars (left) and DM (right) from Merger~I of \mf.  The stars are clustered around $x \sim 10$~kpc along the midplane.  Their spatial distribution is distinct from that of the DM, which is more uniformly distributed although still clustered in the midplane.  The fact that the stars and DM have different spatial distributions results in large local variations in their kinematic distributions.  The top panel of \Fig{fig:ratio_merger} shows how the ratio of the stellar to DM velocity distributions varies across the solar circle.\footnote{In a few of the locations, the most significant merger of \mf~is not Merger~I from \Tab{tab:progenitors}, but rather Merger II.}   On average, the ratio of the stellar and DM distributions is unity, but the spread is quite large---reaching discrepancies of $\gtrsim 2$ in certain locations.  The discrepancies are particularly pronounced in the speed distribution. 

As time proceeds ($t > t_\mathrm{dyn}$), the velocity dispersion of any individual stream decreases as the stars spread out in position space following Liouville's theorem~\citep{Helmi:1999ks}.
 Debris flow~\citep{Lisanti:2011as, Kuhlen:2012fz, Lisanti:2014dva} consists of multiple wraps of these streams, as well as any shells that formed in the process of satellite disruption.  While these contributions are individually cold, their sum is dynamically hot.\footnote{We also note that debris flow may arise from more than one disrupted satellite if the two happened to be on similar orbits and were accreted at comparable times.}  Debris flow is therefore the intermediate state of tidal debris before it becomes fully mixed with the host halo at $t\gg t_\mathrm{dyn}$.  It is identified as kinematic substructure that is coherent over large spatial regions.  

Merger~II of \mi, whose velocity distributions are provided in the bottom panel of \Fig{fig:streamdebris}, is an example of debris flow.  The stellar material from this satellite was accreted at $\zacc \sim2$ and is therefore older than Merger~I of \mf.  In this case, the DM and stars trace each other closely in all velocity components.  The deviations between the distributions are typically under 15$\%$ in each bin, reaching $\sim 30\%$ in some bins along the tails.  Additionally, the DM and stellar debris from this merger are spatially uniform within the solar circle, as shown in the bottom panel of \Fig{fig:position_merger}.  

The velocity distribution of the stars and DM of \mi's Merger~II retain important features that correspond to the satellite's orbital properties, even if the sharp coherence in speed is lost.  For example, the radial velocity distribution is extended and box--like, a feature of satellites on radial orbits.  In such cases, most of the debris is stripped as the satellite moves towards/away from the galactic center, resulting in two peaks of the same radial speed, but opposite direction ($\pm v_r$).   If the dispersion of these peaks is considerably larger than $v_r$, then they bleed into each other, forming a box-like distribution.  This is expected if the turning points of the orbit do not fall near or within the solar circle, so one is primarily sampling material that is removed  while the satellite is on a radial trajectory.

Because the spatial variation of the DM and stars is uniform in this case, their velocity distributions are consistent across localized regions of the solar circle.  The bottom panel of \Fig{fig:ratio_merger} shows the ratio of DM to stellar velocity distributions for this merger. In this case, the ratio is tightly centered about unity over all the regions sampled.  

While we only discussed Merger I of \mf~and Merger~II of \mi~in this subsection, the  conclusions remain unchanged when studying the other significant mergers in both hosts.  The DM and stellar velocity distributions for these mergers are provided in \Fig{fig:other_mergers} and \Fig{fig:other_mergers_mf} of the Appendix.

\section{The Total Dark Matter Distribution}
\label{sec:totaldarkmatter}

In the previous section, we saw that the kinematics of the DM and stars accreted from luminous satellites are well-correlated for older mergers---specifically, the relaxed component and debris flow.  In this section, we will describe how to combine the separate contributions from these populations with the goal of constructing the DM speed distribution at the solar circle.  Sec.~\ref{sec:luminous} will focus on summing the contributions from the relaxed DM with that originating from Mergers I and II in \mi.  As we will see in Sec.~\ref{sec:milky_way}, this methodology will have important applications for the Milky Way, given its similarities to \mi.  Sec.~\ref{sec:untracked} will discuss the `Dark/Unresolved' DM component.

\subsection{Component from Luminous Satellites}
\label{sec:luminous}
  
Taking \mi~as an example, let us consider the scenario where the local stellar halo is dominated by two large mergers (\emph{e.g.}, Merger~I and II) in addition to a relaxed stellar component.  The speed distributions for each of these stellar populations is $f_\mathrm{I}(v)$, $f_\mathrm{II}(v)$, and $f_\mathrm{r}(v)$, respectively, with each normalized to unity.  
The total stellar distribution is therefore given by
\begin{equation} 
f_\mathrm{stellar}(v) =  \xi_{*, \rm{r}} \, f_\mathrm{r} (v) + 
 \xi_{*, \rm{I}} \,  f_\mathrm{ I} ( v) + 
\xi_{*, \rm{II}}  \, f_\mathrm{II} (v)  \, ,
\label{eq:stellarweights}
\end{equation}
where the $\xi_{*}$'s are the observed stellar mass fractions for the components and $\xi_\mathrm{*, r} + \xi_\mathrm{*, I} + \xi_\mathrm{*, II} = 1$.  These values are provided in the first row of Table~\ref{tab:solarcircle}.  Note that we have renormalized the values under the assumption that \emph{all} of the accreted stars belong to either Merger I, II, or the relaxed population, to simplify the discussion.

\begin{figure*}[tb] 
   \centering
	\includegraphics[width=0.99\textwidth]{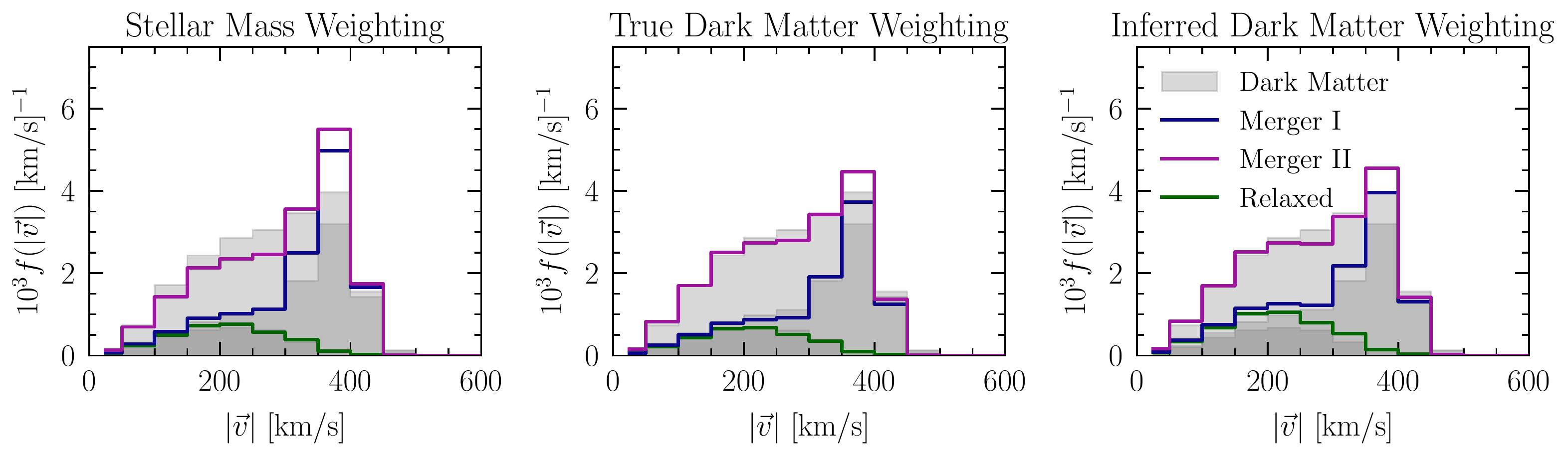}
   \caption{Reconstructing the speed distribution of dark matter from the accreted stars of \mi. The true dark matter distributions for the relaxed component and from Mergers~I and~II are stacked from bottom to top in gray. The distributions inferred from the corresponding  stellar populations are shown by the colored lines (green, blue, and purple, respectively).  To add the stellar speed distributions, we (left) use the stellar mass fractions as per \Eq{eq:stellarweights}; (middle) follow \Eq{eq:exactweights} and take the exact values of the dark matter mass fractions; (right) follow \Eq{eq:normalizations} and take the inferred values of $c_i/c_r$ from the mass-to-light ratios. A similar plot for \mf~is provided in the Appendix as \Fig{fig:scaling_relation_f}.
}
   \label{fig:DMtotal}
\end{figure*}

The left-most panel of \Fig{fig:DMtotal} shows the stacked speed distributions for the stars associated with the relaxed component (green solid), Merger~I (blue solid), and Merger~II (purple solid), combined according to \Eq{eq:stellarweights}.  This corresponds to the total speed distribution for the accreted stars.  Let us compare this to the stacked distributions for the DM associated with these same populations (shown in gray).  Clearly, the two do not match.  We have already seen that the stellar distributions for the separate populations of \mi~reproduce those of the DM (see \Fig{fig:virialized}, \Fig{fig:streamdebris}, and \Fig{fig:other_mergers}).  Therefore, the source of the discrepancy arises from using the stellar mass fractions in \Eq{eq:stellarweights}. 

To reproduce the total DM distribution, we should instead use the DM mass fraction $\xi_\mathrm{dm}$ for each component as its appropriate weight in the sum:
\begin{equation} 
f_\mathrm{dm}(v) =  \xi_\mathrm{dm, r} \, f_\mathrm{r} (v) + 
\xi_\mathrm{dm, I} \,  f_\mathrm{ I} ( v) + 
\xi_\mathrm{dm, II} \, f_\mathrm{II} (v)  \, .
\label{eq:exactweights}
\end{equation}
The $\xi_\mathrm{dm}$ values are provided in the second row of Table~\ref{tab:solarcircle}.  Using these exact weights, we can stack the stellar distributions according to \Eq{eq:exactweights}; the result is shown in the middle panel of \Fig{fig:DMtotal} and reproduces the total DM distribution, as desired.

In reality, we do not know the exact DM mass fraction of each component, so we need a way to infer its value.  To do so, it will be useful to recast \Eq{eq:exactweights} as follows: 
\begin{equation} \label{eq:normalizations}
f_\mathrm{dm}(v) =  N \left( \xi_{*, \rm{r}} \, f_\mathrm{r} (v) + \frac{ c_{\scriptscriptstyle{\mathrm{I}}}  }{c_\mathrm{r}} \, \xi_{*, \rm{I}} \,  f_\mathrm{I} ( v) + \frac{c_{\scriptscriptstyle{\mathrm{II}}}}{c_\mathrm{r}} \,  \xi_{*, \rm{II}}  \, f_\mathrm{II} (v) \right) \, ,
\end{equation}
where $N$ is a normalization constant, and $c = M_\mathrm{DM} / M_\mathrm{*} $ for each population.  The value of $c$ tells us about the relative amount of DM and stars that each merger leaves at the solar circle.  The DM-stellar mass fractions are provided in the third row of Table~\ref{tab:solarcircle}, and the true values of $c_{\scriptscriptstyle{\mathrm{I(II)}}} /c_\mathrm{r}$ are provided in the sixth row.

To approximate the value of $M_\mathrm{DM} / M_\mathrm{*}$ for a  given merger, we will use its mass-to-light ratio.  That is, we will assume that $c \approx M_\mathrm{peak} / M_\mathrm{*, total} $.
Note that the relaxed population is itself the sum of several mergers.  
Moving forward, we treat these old mergers as a single population with some average metallicity and $M_\mathrm{peak}/M_\mathrm{*, total}$.  

At first glance, this may seem like a poor approximation as the true $M_\mathrm{peak}/M_\mathrm{*, total}$ ratio (fourth row of Table~\ref{tab:solarcircle}) is approximately an order of magnitude larger than the corresponding $M_\mathrm{DM}/M_\mathrm{*}$ ratio.  However, the reduction between the two ratios is roughly consistent between the separate populations, and thus cancels out when taking $c_{\scriptscriptstyle{\mathrm{I(II)}}} /c_\mathrm{r}$.  We therefore conclude that $c \approx M_\mathrm{peak} / M_\mathrm{*, total}$ is an adequate approximation so long as each satellite loses roughly the same fraction of DM from its halo before it reaches the solar circle.

\begin{table*}[t]
\centering
\footnotesize
\renewcommand{\arraystretch}{1.5}
\begin{tabular}{C{4 cm} | C{1.4cm}  C{1.4cm} C{1.4cm}  | C{1.4cm}  C{1.4cm} C{1.4cm}}
\Xhline{3\arrayrulewidth}
&\multicolumn{3}{c|}{\textbf{\texttt{FIRE m12i} Host Halo}} &  \multicolumn{3}{c}{\textbf{\texttt{FIRE m12f} Host Halo}} \\
&  Relaxed & I & II & Relaxed &I & II    \\ 
 \hline
Stellar Fraction at Solar Circle& 0.17 & 0.49 & 0.34  & 0.13 & 0.45 & 0.41  \\
Dark Matter Mass Fraction & 0.18 & 0.35 & 0.47 & 0.17 & 0.34 & 0.49  \\ 
$M_\mathrm{DM}/M_\mathrm{*}$ at Solar Circle&  30 &19 & 35 &  14 & 6 & 12  \\
True $M_\mathrm{peak}/M_\mathrm{*,total}$ &  523 & 122 &  101 & 562 & 82 & 66 \\
Inferred $M_\mathrm{peak}/M_\mathrm{*,total}$ &  239 & 135 &  192 & 176 & 54 & 71 \\
True $c_i /c_\mathrm{r}$ & --- & 0.6 & 1.2   & --- & 0.4 & 0.8  \\
Inferred $c_i /c_\mathrm{r}$ & --- & 0.6 & 0.8  & --- & 0.3 & 0.4  \\
    \Xhline{3\arrayrulewidth}
  \end{tabular}
  \caption{Relevant fractions at the solar circle for the~\mi~and~\mf~host halos, divided by the relaxed population and Mergers I--II.  Note that Merger~III is included in the relaxed component.  From top to bottom, we provide the following: (i) the stellar mass from each component at the solar circle assuming only the relaxed component and Mergers I--II; (ii) the dark matter mass from each component, relative to the total accreted dark matter mass at the solar circle from the relaxed component and Mergers I--II; (iii) the dark matter mass from each component, relative to its stellar mass at the solar circle; 
  (iv) the true $M_\mathrm{peak}/ M_\mathrm{*,total}$ from the simulation; (v) the inferred $M_\mathrm{peak}/ M_\mathrm{*,total}$ from the procedure described in the text; (vi) the true $c_i/c_\mathrm{r}$ ($i =$~I or II) values; (vii) the inferred $c_i/c_\mathrm{r}$ values using the estimated mass-to-light ratio. }
  \label{tab:solarcircle}
  \end{table*}

\begin{figure*}[tb] 
   \centering
	\includegraphics[width=0.99\textwidth]{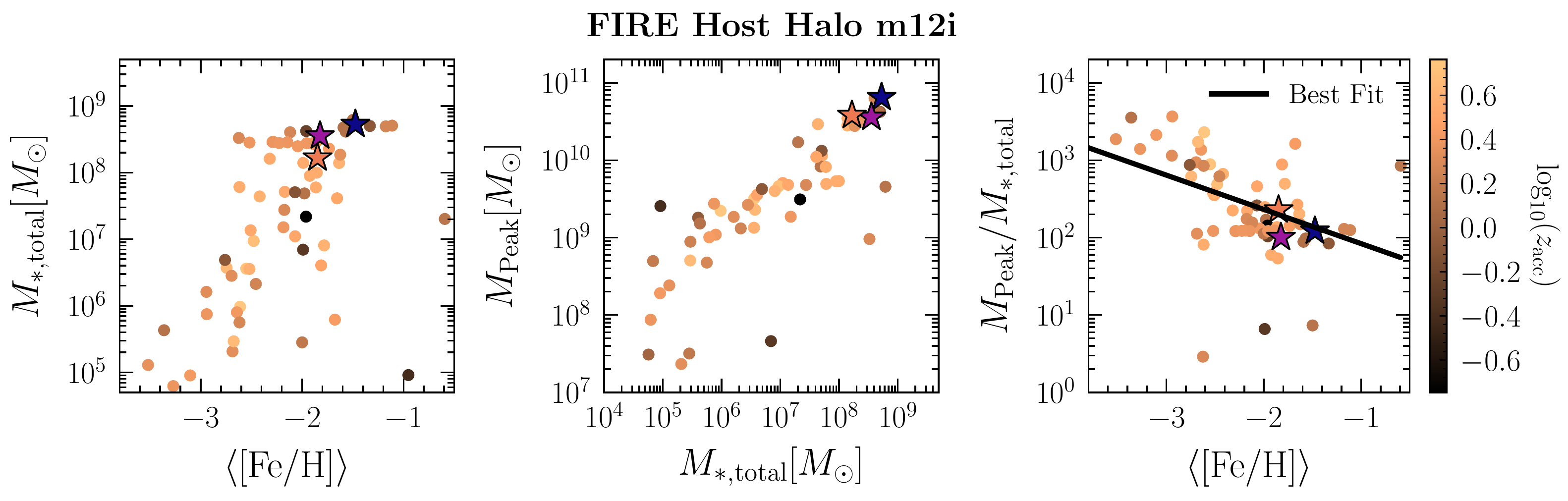}
   \caption{(Left) The relation of stellar mass and metallicity for the subhalos in \mi~that contribute stars within the solar circle. (Middle) The relation of peak halo mass and stellar mass for the same subhalos.  (Right) The ratio of peak halo mass to stellar mass as a function of the average metallicity of each subhalo.  The best-fit line, defined in \Eq{eq:scaling_relation}, is shown in solid black.  In each panel, the stars correspond to Mergers I--III; their color convention matches that of Fig.~\ref{fig:origins}.  The color of the points corresponds to the average accretion redshift for the stars in the merger.  
}
   \label{fig:scaling_relation}
\end{figure*}

To extrapolate the mass-to-light ratio, we use the present-day stellar mass-metallicity ($M_\mathrm{*, total} - \langle \FeH \rangle $) and peak halo mass--stellar mass  ($M_\mathrm{peak}-M_\mathrm{*, total}$) relations.  We now demonstrate this within the context of \mi, saving a discussion of the Milky Way application to Sec.~\ref{sec:milky_way}. 
The left and middle panels of \Fig{fig:scaling_relation} show the $M_\mathrm{*, total} - \FeH$ and $M_\mathrm{peak}-M_\mathrm{*, total}$ relations for \mi.\footnote{Note that \Fig{fig:scaling_relation} only includes the progenitor subhalos that eventually contribute debris within the solar circle.  However, the corresponding relations for the Milky Way are provided for all observed dwarf galaxies at redshift $z=0$.}   Taken together, these can be used to obtain the dependence of the $M_{\rm{peak}}/M_\mathrm{*, total}$ ratio on $\langle \FeH\rangle$, which is provided in the right panel of \Fig{fig:scaling_relation}.   The mass-to-light ratio $M_{\rm{peak}}/M_\mathrm{*, total}$ is inversely proportional to the metallicity, with the more DM-dominated galaxies typically associated with more metal-poor stars.  The approximately linear relationship is well-fit by
\be 
\log_{10} \left( \frac{M_{\rm{peak}}}{M_\mathrm{*, total}} \right) = 1.48 - 0.44 \,\langle \rm{ [Fe/H] } \rangle \, ,
\label{eq:scaling_relation}
\ee
 indicated by the solid black line in \Fig{fig:scaling_relation} (right). Given the average metallicities for Mergers I--II in \mi, we infer that  $M_{\rm{peak}}/M_\mathrm{*, total} = \{ 135, 192 \}$, respectively, which are $\mathcal{O}(1)$ of the true values $\{122, 101 \}$. 
The slight offset is evident in \Fig{fig:scaling_relation} (right) where Mergers I--II are denoted by the colored stars and fall slightly above/below the best-fit line.  Similarly, we estimate that the relaxed population\footnote{There are many ways to compute the mean of $M_\mathrm{peak}/ M_\mathrm{*,total}$ of the relaxed population. In Table \ref{tab:solarcircle}, we present the values of the mean over all relaxed subhalos, however these values might be artificially high. If one were to weigh the average by the subhalo mass for example, the value for \mi (\mf) would drop to $329$ ($213$). } is comprised of mergers with $\langle M_\mathrm{peak}/M_\mathrm{*, total} \rangle= 239$ given that their average metallicity is $\langle \FeH \rangle = -2.04$.  

Given an inferred $M_{\rm{peak}}/M_\mathrm{*, total}$ for each stellar component, we can estimate $c_i/c_r$ ($i = $~I, II).  The values for Mergers I--II of \mi~are provided in the seventh row of \Tab{tab:solarcircle}, and they compare well to the true values.  Using these weights in \Eq{eq:normalizations}, the distribution inferred from the stars is an excellent approximation of the underlying DM distribution, even if not an exact reproduction.  The final result is shown in the right panel of \Fig{fig:DMtotal}.   

We apply the same procedure to \mf~and provide the corresponding figure in the Appendix as \Fig{fig:scaling_relation_f}.  In this case, the inferred values of $c_i/c_r$ are close to their true values (see Table~\ref{tab:solarcircle}) but the stellar distributions do not do a good job reconstructing the total DM.  The failure is due to the discrepancy in the DM and stellar speed distribution for Merger~I (a stream), which we discussed in Sec.~\ref{sec:debris}.  

\subsection{Untracked Component}
\label{sec:untracked}

Next, we consider the DM in the `Dark/Unresolved' component.  As already discussed, this component consists of DM that originates from subhalos whose galaxies are not adequately resolved, truly dark subhalos, unresolved subhalos, or smooth accretion.  In the first case, the component may actually be tracked by stars.  For the other cases, we do not expect stars to be brought in along with the DM.  Because we cannot further distinguish between these separate contributions, we conservatively group them together and study their total velocity distribution.  

\begin{figure*}[tb] 
   \centering
	\includegraphics[width=7in]{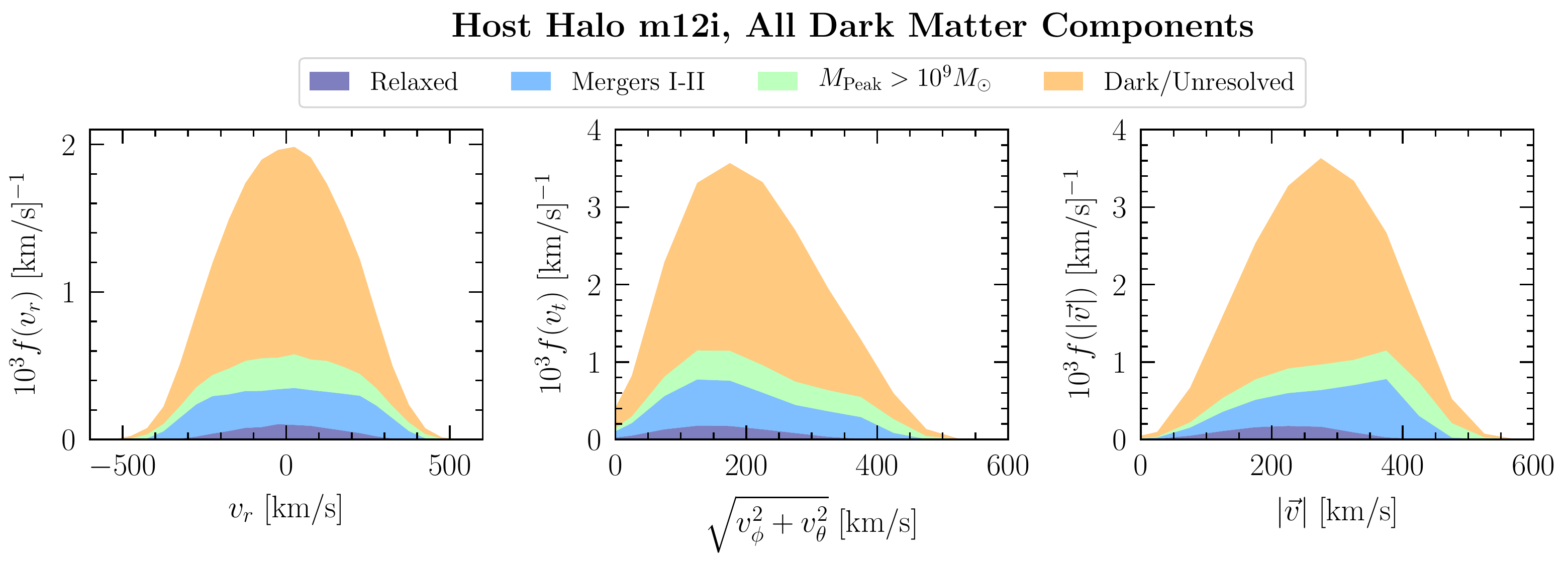}
   \caption{Present-day velocity distributions for all dark matter in the solar circle of \mi.  The contributions are divided by origin: the relaxed component (purple), Mergers I~and~II (blue), all other mergers from subhalos with $M_\mathrm{peak} > 10^9$~M$_\odot$ (cyan), and the `Dark/Unresolved' component (orange). The equivalent plot for \mf~is provided as \Fig{fig:unvirializedmf} in the Appendix.  }
   \label{fig:unvirializedmi}
\end{figure*}

\Fig{fig:unvirializedmi} plots the radial, tangential, and speed distributions for the `Dark/Unresolved' component of \mi.  The distributions are stacked on top of the distributions for the relaxed population and Mergers I--II.  We also include the contribution from DM that originates from sub-dominant mergers with $M_\mathrm{peak} > 10^9$~M$_\odot$; this contribution is similar to that of Mergers I--II.  The additional DM from the `Dark/Unresolved' component has two important effects.  First, it decreases the overall dispersion in the radial velocity, smoothening out the kinematic structure left behind by the recent mergers.  Second, it shifts the peak in the speed distribution to a value that lies closer (but still above) that of the relaxed component.  As we see from \Fig{fig:origins}, the `Dark/Unresolved' contribution enters the solar circle at redshift $\zacc \lesssim 2$, which explains why its overall speed is faster, on average, than that of the relaxed component.  

We emphasize that it is not possible to infer the fraction of DM originating from smooth accretion and/or dark subhalos in the Milky Way directly from simulations.  The primary challenge is that both depend sensitively on the accretion history of the simulated host halo, which may not replicate that of the Milky Way.  
The wide halo-to-halo variation has already been underscored by a separate study of ten~\texttt{Aquarius} halos~\citep{2011MNRAS.413.1373W}, which found large variations in the fractional contribution of each population between different Milky Way realizations.  It is therefore imperative to develop methods of characterizing the DM contribution from smooth accretion and dark subhalos empirically.  This requires its own dedicated study.
 
\section{The Local Dark Matter Distribution in the Milky Way}
\label{sec:milky_way}

We now apply the formalism developed in Sec.~\ref{sec:correlation} and~\ref{sec:totaldarkmatter} to our own Galaxy with the aim of inferring the local DM speed distribution from observations.  \cite{necib2018} characterized the velocity distribution of the local accreted stellar population using a cross-match of \emph{Gaia}~DR2 data \citep{2016A&A...595A...4L,2018arXiv180409365G} and SDSS~\citep{2012ApJS..203...21A}.  They characterized a metal-poor `halo' population with average metallicity $\langle \FeH \rangle_{\rm{halo}} = -1.82$ that is nearly isotropic and comprises $\sim 24\%$ of the local accreted stars within heliocentric distances of 4~kpc and above $|z| > 2.5$~kpc of the midplane.\footnote{Note that the volume of study in \cite{necib2018} is outside the solar circle, as defined in this work.}  It is the parallel of the relaxed population discussed in Sec.~\ref{sec:virialized}.  The Milky Way's relaxed component constitutes a larger fraction of the stellar halo and is moderately more metal-rich than that of \mi~or \mf.

Additionally, the authors characterized the kinematics of a younger stellar population with average metallicity $\langle \FeH \rangle_{\rm{subs}} = -1.39$.  This substructure, referred to as the \emph{Gaia} Sausage or \emph{Gaia} Enceladus, is an example of debris flow.  Like Merger~II of \mi, its velocity distribution is  highly radial and spatially uniform within the SDSS footprint.  However, it contributes a much larger fraction of the local accreted stars ($\sim 76 \%$) than does Merger~II of~\mi~($\sim 30\%$).  

As the inner Milky Way appears to be dominated by the stellar debris of one single large merger, its composition is simpler than that of either \mi~or \mf.  Consequently, we need only consider the sum of two terms when building the distribution of local DM speeds in the Galaxy:
\begin{equation}\label{eq:MWratio}
f_\mathrm{dm}(v) = N \left( \xi_\mathrm{*, halo} \, f_\mathrm{halo}(v) + \frac{c_\mathrm{subs}}{c_\mathrm{halo}} \, \xi_\mathrm{*, subs} \, f_\mathrm{subs}(v) \right)\, ,
\end{equation}
where the first term corresponds to the relaxed component and the second term corresponds to the substructure.  Note that we identify these contributions with the terms `halo' and `subs' as in~\cite{necib2018}.  The ratio $c_\mathrm{subs}/c_\mathrm{halo}$ can be determined following the procedure outlined in  \Sec{sec:luminous}, but using relations specific to the Milky Way.

We adopt the $M_\mathrm{*, total}-\FeH$ relation from \cite{Kirby:2013wna}:
\be
\langle \rm{[Fe/H]} \rangle=(-1.69 \pm 0.04)+(0.30 \pm 0.02) \log_{10} \left(\frac{M_\mathrm{*, total}}{10^6 \, M_\odot} \right) \, ,
\label{eq: FeH}
\ee
which applies to dwarf galaxies of the Milky Way at redshift $z=0$.  The root-mean-square about the best-fit line is 0.17~dex.  This linear relation holds over many orders of magnitude in stellar mass, from $M_\mathrm{*, total} \sim 10^4$--$10^9$~M$_\odot$.  Data from SDSS suggest that the trend roughly continues up to $M_\mathrm{*, total} \sim 10^{12}$~M$_\odot$~\citep{Gallazzi:2005df}.  \Eq{eq: FeH} is similar to the $M_\mathrm{*, total}-\FeH$ relation recovered in the \textsc{Fire-2} simulations (see \emph{e.g.}, \Fig{fig:scaling_relation}).  However, while the simulations reproduce the observed slope, they find systematically lower values of iron abundance~\citep{Escala2017}.  This offset is likely due to specific choices made in the modeling of the delay time distribution and yields of Type~Ia Sne.

The \cite{Kirby:2013wna} relation applies to observed dwarf galaxies at redshift $z=0$, while the desired quantity is the stellar mass of galaxies disrupted at earlier redshifts.  In this work, we assume that there is no redshift dependence to the stellar mass-metallicity relation.  To estimate the size of this dependence, we can combine \Eq{eq: FeH} with the redshift evolution inferred from simulations.  Taking as an example the work of~\cite{Ma:2015ota}, we assume a shift in average metallicity that goes as $\Delta \FeH = 0.67 \left[ (\exp(-0.5 z) -1 \right]$.  For a merger at redshift $z = 1$, this leads to $\Delta \FeH = - 0. 26$.  A merger at redshift $z=3$, is associated with a shift of $-0.52$.  This correction shifts the expected metallicity down by some constant at any given redshift.  In our case, though, we are only interested in the relative difference in metallicities between the substructure and halo populations---and this does not change with redshift evolution.  As a result, $c_\text{subs}/c_\text{halo}$ is unaffected.

To estimate the peak halo mass, we follow the same procedure outlined by \cite{Garrison-Kimmel:2016szj}.
Above $M_{\rm{peak}} \gtrsim 10^{11.5}$~M$_\odot$, this $M_\mathrm{peak}-M_\mathrm{*, total}$ relation maps onto that of \cite{Behroozi:2012iw}, which has a constant log-normal scatter of $\sigma = 0.2$~dex about the median value of $M_\mathrm{*, total}$.  For lower-mass galaxies with $M_{\rm{peak}} \lesssim 10^{11.5}$~M$_\odot$, the stellar mass is effectively a power law in peak halo mass.  Specifically, $M_{*} \propto M_{\rm{peak}}^\alpha$ where the slope $\alpha$ depends on the assumed log-normal scatter, $\sigma_v$, about the mean value of $M_\mathrm{*, total}$.  We use the growing-scatter model of~\cite{Garrison-Kimmel:2016szj} where the value of $\sigma_v$ is allowed to grow linearly as $\log_{10} M_\mathrm{peak}$ decreases.  That is,  
\begin{equation}
\sigma_v = 0.2 + v \times (\log_{10} M_\mathrm{peak} - \log_{10} M_1) \, ,
\label{eq: sigmav}
\end{equation} 
where $M_1 \sim 10^{11.5}$~M$_\odot$ and $v$ sets how the scatter increases.  The best-fit power-law slope in this case is 
\begin{equation}
\alpha \simeq 0.25 \, v^2 - 1.37 \, v + 1.69 \, .
\label{eq: alpha}
\end{equation} 
We take $v= -0.1$ as our benchmark value.

\begin{figure}[tb] 
   \centering
	\includegraphics[width=0.48\textwidth]{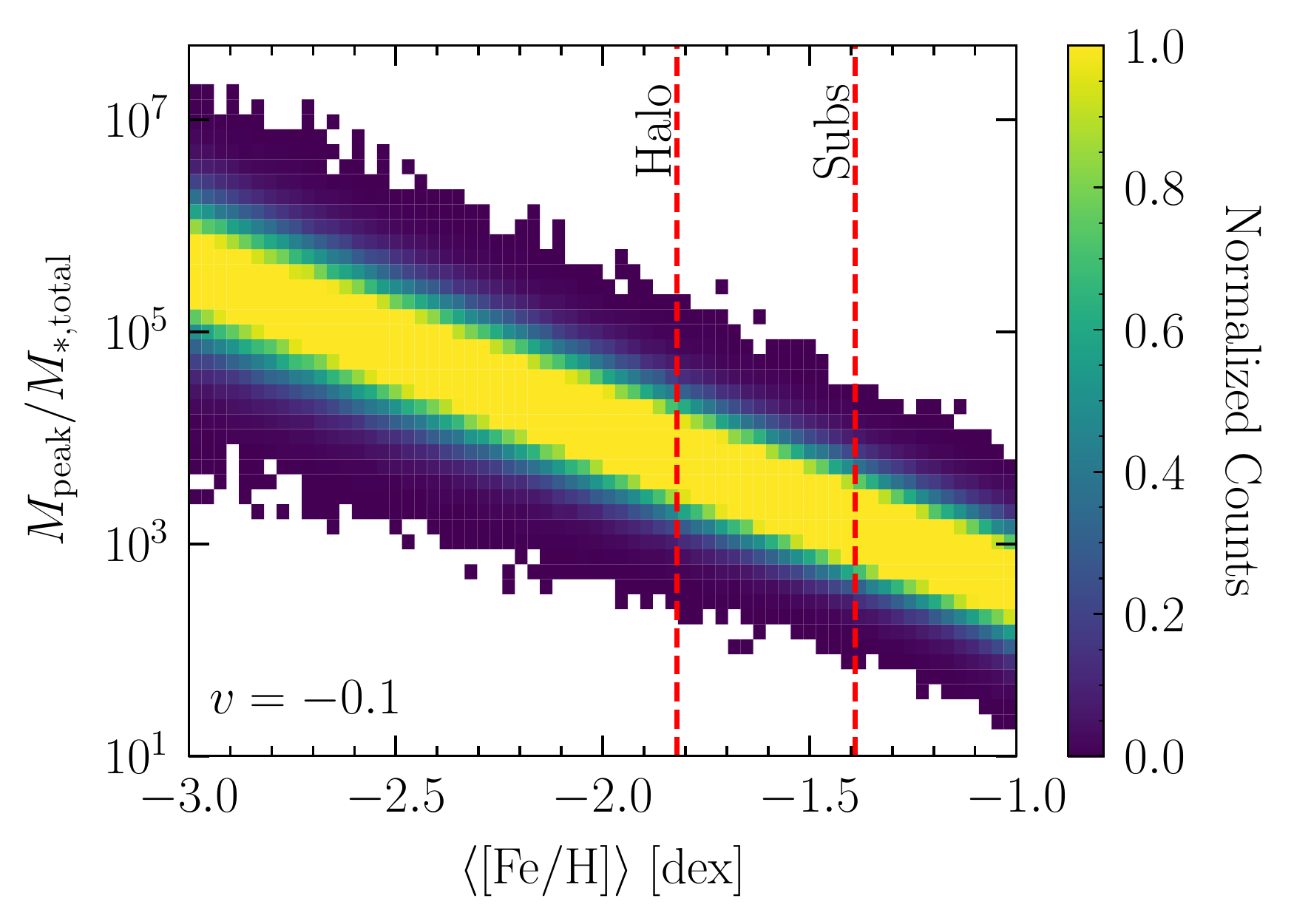}
   \caption{Estimated $M_{\rm{peak}}/M_\mathrm{*, total} - \langle \FeH \rangle$ relation, assuming the growing-scatter model of  \cite{Garrison-Kimmel:2016szj} with $v = -0.1$. 
 The average metallicities of the halo and substructure components in the Milky Way, as derived in~\cite{necib2018} are indicated by the red dashed lines.  }
   \label{fig:observation}
\end{figure}

We note that this $M_\mathrm{*, total}-M_{\rm{peak}}$ relation was derived for DM-only simulations and that the presence of a baryonic disk can have important effects.  The expectation is that the disk will tidally destroy infalling subhalos, requiring that the predicted $M_\mathrm{*, total}$ (for given $M_\text{peak}$) must be shifted to higher values in order to recover the Milky Way's cumulative stellar mass function \citep{Garrison-Kimmel:2017zes}.  This, in turn, would result in a more shallow power-law fall off. 

We perform a Monte Carlo procedure to estimate the relative amount of local DM in substructure as opposed to the halo population (\emph{e.g.}, $c_{\rm{subs}}/c_{\rm{halo}}$).  The procedure is as follows: 
\begin{enumerate}
\item We use the  $M_\mathrm{peak}-M_\mathrm{*, total}$ relation to estimate the associated stellar mass, for a given $M_{\rm{peak}}$.  The value of $M_\mathrm{*, total}$ is randomly selected from a normal distribution with mean given by the growing-scatter model of~\cite{Garrison-Kimmel:2016szj}, with self-consistent $v$, $\sigma_v$, and $\alpha$ from \Eq{eq: sigmav} and \Eq{eq: alpha}.  This yields a prediction for the $M_{\rm{peak}}/M_\mathrm{*, total}$ ratio.  We demand that $M_\mathrm{peak} > 5 \, M_\mathrm{*, total}$. 
\item Using this stellar mass, we estimate the metallicity by randomly selecting  $\langle \FeH \rangle$ from a normal distribution with mean given by \Eq{eq: FeH} and dispersion of $\sim 0.17$~dex.  
\item We repeat the previous two steps 500 times to build a distribution of $M_{\rm{peak}}/M_\mathrm{*, total}$ versus $\langle \FeH \rangle$.  The result is shown in \Fig{fig:observation}.
\item We randomly select a point with metallicity $\langle \FeH \rangle \sim -1.39$, as per the substructure population, and another with metallicity $\langle \FeH \rangle \sim -1.82$, as per the halo population.  The ratio of their respective $M_{\rm{peak}}/M_\mathrm{*, total}$ values yields the $c_{\rm{subs}}/c_{\rm{halo}}$ weighting factor.  Repeating this $8\times 10^6$ times allows us to quantify the 16-50-84$^\text{th}$ percentiles of this factor.  
\end{enumerate}
For the $v = -0.1$ benchmark, we find that  
\be
\frac{c_{\rm{subs}}}{c_{\rm{halo}}} = 0.23^{+0.43}_{-0.15} \, .
\label{eq:SausRatio}
\ee 
Substituting this back into \Eq{eq:MWratio}, we find that $42 ^{+26}_{-22}\%$ of the local DM that originates from luminous satellites is in debris flow.\footnote{To simplify this calculation, we did not convolve the error on the stellar fraction from the best fit in \cite{necib2018}. We expect it to be subdominant to the error from \Eq{eq:SausRatio}.  }  This value is consistent, within the range of uncertainty, with values estimated using kinematic arguments in~\cite{Evans2018}.

One might notice that \Eq{eq:SausRatio} is systematically lower than the reweighting factors found for $\mi$.  This is because there is a greater difference in metallicity between the relaxed and substructure populations in \mi, compared to what is observed in the Milky Way.  Because the halo and substructure populations in the Milky Way are closer to each other in average metallicity, the amount of DM that each contributes is commensurate between the two.

\begin{figure}[tb] 
   \centering
	\includegraphics[width=0.45\textwidth]{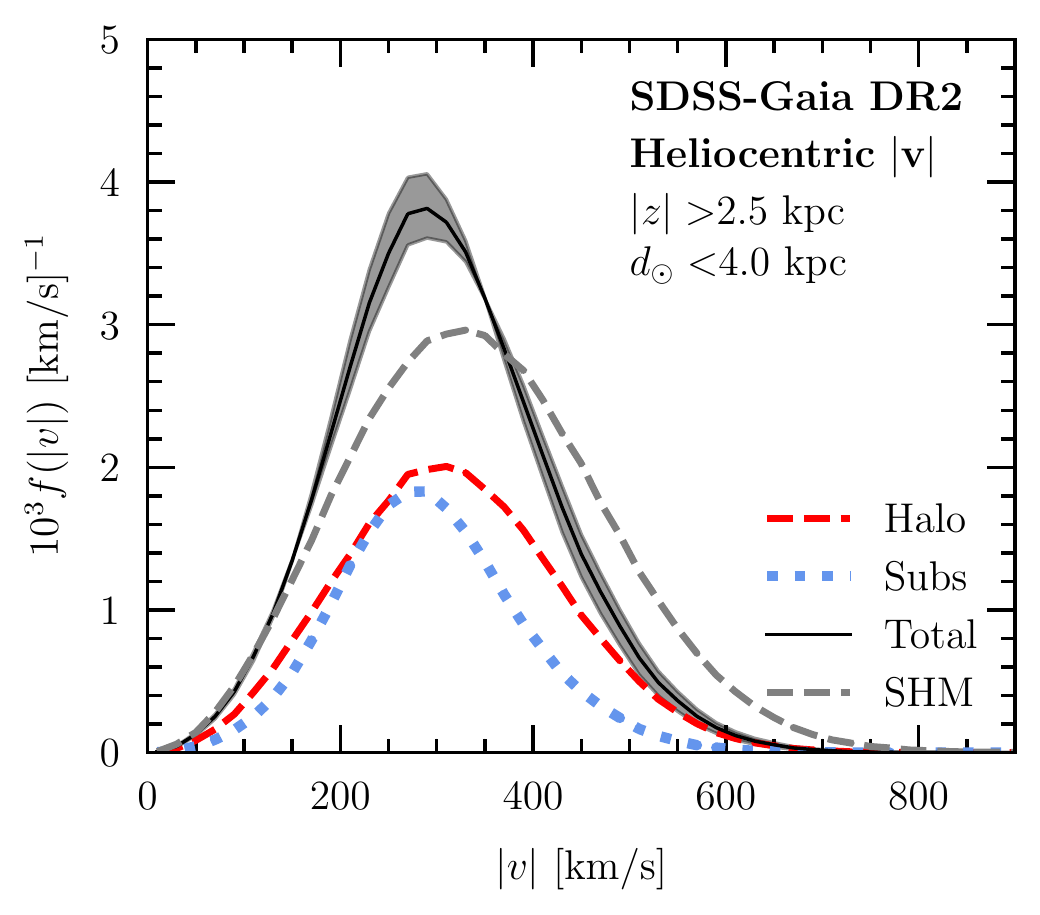}
   \caption{Updated heliocentric velocity distribution from \cite{necib2018}. This distribution takes into account the relative dark matter contribution between the substructure and the halo component, as given by \Eq{eq:SausRatio}. This distribution applies to the dark matter accreted from luminous satellites, and does not account for potential contributions from dark subhalos or smooth accretion.  Data for this distribution is publicly available at \url{https://linoush.github.
io/DM_Velocity_Distribution/}.}
   \label{fig:helio_vel}
\end{figure}

\Fig{fig:helio_vel} shows the heliocentric velocity distribution inferred from the SDSS-\emph{Gaia}~DR2 data.  The halo and substructure distributions (red dashed and blue dotted, respectively) were derived in~\cite{necib2018}.  When summing their contributions (black solid), the relative fraction is set by \Eq{eq:MWratio} and \Eq{eq:SausRatio}.  The gray band denotes the uncertainty from the inferred value of $c_{\rm{subs}}/c_{\rm{halo}}$.  For comparison, we also show the Standard Halo Model (gray dashed), assuming a Maxwell-Boltzmann distribution with a dispersion $\sigma = 220/\sqrt{2}$~km/s.

\section{Conclusions}
\label{sec:conclusions}

In this paper, we studied two cosmological zoom-in hydrodynamic simulations of Milky Way-mass galaxies from the \emph{Latte} suite of \textsc{Fire}-2 simulations.  Our primary goal was to understand how the DM and stars accreted from luminous satellite galaxies trace each other in the inner regions of a Milky Way--like galaxy.  In each of these host galaxies, we focused on the accreted material in the solar circle (defined as $|r-r_{\odot}| < 2$ kpc  and $|z| \leq 1.5$~kpc with $r_\odot$ the solar radius), which is most relevant for ground-based DM direct detection experiments. 

The accreted DM and stars in the solar circle can be divided into three separate components whose general behavior is summarized as follows: 
\begin{itemize}
\item  The `relaxed' DM and stellar component is accreted from the oldest mergers ($\zacc \gtrsim 3$).  At these early times, the proto-galaxy is still evolving and changes to the galactic potential redistribute the energies of the DM and stellar debris, mixing them fully in phase space.  As a result, the present-day velocity distributions of the DM and stars from these oldest mergers are well-correlated.   The metal-poor sub-component of local stars is an adequate proxy for the relaxed population.  We find that stars with metallicity $\FeH \lesssim -2$ to $-3$ trace the relaxed distributions reasonably well in the \emph{Latte} hosts, consistent with previous results from~\cite{Herzog-Arbeitman:2017fte}.  It is possible that the low-metallicity sample may be contaminated by stars from more recent mergers whose metallicity distributions have low-metallicity tails. Statistical clustering algorithms---such as that used in~\cite{necib2018}---can ameliorate such contamination.
\item Once the proto-galaxy is in place, smaller mergers continue to the present day.  The tidal debris from these mergers evolves in phase space following Liouville's theorem.  As a satellite falls into the galaxy, it leaves behind a trail of tidal debris.  If the time since infall is relatively short, then this material is typically in a stream, and is clustered in both position and velocity space.  In such cases, we find that there can be significant spatial variations in the DM and stars, which lead to discrepancies in their velocity distributions.  
\item If the time since infall is longer and the satellite has completed multiple orbital wraps, then the spatial distribution of its tidal debris is well-mixed, but the kinematic substructure is still preserved.  This class of substructure is referred to as debris flow.  We find that the velocity and spatial distributions of the DM and stars from these mergers are well-correlated.  Unlike the case of stellar streams, the distributions do not exhibit large local variations.
\end{itemize}
As we have demonstrated with the \emph{Latte} \textsc{Fire-2} simulations, the DM-stellar correlations are robust for both the relaxed and debris flow populations, and hold despite the significant differences in the merger histories of the two host halos studied here.   The conclusions are specific to the solar circle, where our study is focused.  For the most significant mergers (Mergers I--III in Table~\ref{tab:progenitors}), we find that much of the halo has been stripped off by the time the satellite has sunk to the solar radius.  As a result, the DM being removed as the satellite passes through the solar circle is the most bound, similar to the stars.  

In the case of streams, care needs to be taken in extrapolating the kinematic DM properties from the stellar distributions due to large localized variations that can arise. Dedicated simulations may be needed to better quantify the expected discrepancies between the DM and stellar debris from a particular merger.  Such simulations may be warranted to study the potential DM contribution from stellar streams, such as S1, in the solar neighborhood~\citep{Myeong:2017skt, 2018arXiv180407050M, OHare:2018trr}. 

The total DM velocity distribution at the solar circle can be built up from the separate components described above if one can infer the relative amounts of DM brought in by each merger.  We provide a simple procedure to do so, which combines the mass-metallicity relation with abundance matching to relate the $M_{\rm{peak}}/M_\mathrm{*, total}$ ratio to the average metallicity $\langle \FeH \rangle$ of a merger. This relation allows us to estimate the relative amounts of DM to stars brought in by each merger.  In this way, we can build the total velocity distribution for the DM associated with luminous mergers. 

The results of our work on the \emph{Latte} hosts is pertinent in light of the recently discovered stellar debris field in the Solar neighborhood~\citep{2018arXiv180203414B, 2018arXiv180606038H}.  These stars can be divided into a metal-poor and nearly isotropic population and a more metal-rich and radially-biased population.  Using a Gaussian clustering algorithm, \cite{necib2018} recently extracted the velocity distributions of these two components using data from the SDSS-\emph{Gaia} cross-match.  The two components correspond to a relaxed stellar population and debris flow and should be well-traced by the DM removed from the same set of mergers, following our study of the \emph{Latte} hosts.  Using the rescaling relations from \Sec{sec:milky_way}, we estimate that $42 ^{+26}_{-22}\%$ of the local DM accreted from luminous satellites is in debris flow.  

The method described in this paper does not, by assumption, account for DM contributions from dark subhalos or smooth accretion, which should not be associated with stars.  In the \emph{Latte} hosts studied here, we are not able to distinguish this contribution from DM arising from unresolved DM (sub)halos or halos whose galaxies are not resolved.  However, the distinguishing power will improve as the stellar and DM mass resolution improves.  In the \emph{Latte} hosts, this DM contribution (which we label as `Dark/Unresolved') comes in at redshifts $z \lesssim 2$, so it has, on average, larger speeds than the older relaxed component.  

It is challenging to extract conclusions regarding dark subhalos or smooth accretion in simulations to our own Galaxy.  Previous studies using high-resolution DM-only $N$-body simulations have found considerable variation in the potential origin of DM in the solar neighborhood.  For example, the DM halo in the \texttt{Via Lactea} simulation is rapidly built up around redshift $z\sim 1.7$ and then remains essentially stationary until present time~\citep{Diemand:2007qr}.  In some \texttt{Aquarius} halos, the DM in the solar neighborhood is nearly all in place before $z\sim6$, whereas in others, most of the DM accreted more recently~\citep{2011MNRAS.413.1373W}.  This variation underscores the importance of studying a variety of simulated halos to better understand how the fraction of local DM from dark subhalos or smooth accretion depends on merger history.  Only in this way can we robustly extrapolate conclusions to the Milky Way.  

Finally, we emphasize that all results regarding the DM-stellar correspondence that we draw from the \textsc{Fire-2} simulations assume cold, collision-less DM.  It will be important to understand how these conclusions generalize to a broader class of DM models where the DM and stellar trajectories may be different, by assumption.  Some classic examples include self-interacting or ultralight scalar DM models.  

For readers who would like to use the empirical velocity distributions from \cite{necib2018} to model the local DM distribution from luminous satellites, we provide interpolated functions at \url{https://linoush.github.io/DM_Velocity_Distribution/}.  The separate contributions from the halo and substructure distributions can be combined following the prescription in Sec.~\ref{sec:milky_way}.  

\break

\section*{Acknowledgements}

We thank V.~Belokurov, E.~Kirby, A.~Peter, and D.~Spergel for useful conversations. This research made use of \texttt{Astropy}~\citep{2013A&A...558A..33A} and \texttt{IPython}~\citep{PER-GRA:2007}. LN is supported by the DOE under Award Number DESC0011632, and the Sherman Fairchild fellowship. ML is supported by the DOE under Award Number DESC0007968 and the Cottrell Scholar Program through the Research Corporation for Science Advancement.  Support for SGK was provided by NASA through Einstein Postdoctoral Fellowship grant number PF5-160136 awarded by the Chandra X-ray Center, which is operated by the Smithsonian Astrophysical Observatory for NASA under contract NAS8-03060.
AW was supported by NASA through ATP grant 80NSSC18K1097 and grants HST-GO-14734 and HST-AR-15057 from STScI.  CAFG was supported by NSF through grants AST-1517491, AST-1715216, and CAREER award AST-1652522, by NASA through grants NNX15AB22G and 17-ATP17-0067, and by a Cottrell Scholar Award from the Research Corporation for Science Advancement. Support for PFH, SGK, and RES was provided by an Alfred P. Sloan Research Fellowship, NSF Collaborative Research Grant \#1715847 and CAREER grant \#1455342, and NASA grants NNX15AT06G, JPL 1589742, 17-ATP17-0214. Numerical calculations were run on the Caltech compute cluster ``Wheeler,'' allocations from XSEDE TG-AST130039 and PRAC NSF.1713353 supported by the NSF, and NASA HEC SMD-16-7592. DK was supported by NSF grant AST-1715101 and the Cottrell Scholar Award from the Research Corporation for Science Advancement.
This work was performed in part at Aspen Center for Physics, which is supported by National Science Foundation grant PHY-1607611.
We used
computational resources from the Extreme Science and
Engineering Discovery Environment (XSEDE), supported by
NSF.

\clearpage
\def\bibsection{} 
\bibliographystyle{aasjournal}
\bibliography{sims}

\begin{thebibliography}{}
\expandafter\ifx\csname natexlab\endcsname\relax\def\natexlab#1{#1}\fi

\bibitem[{{Ahn} {et~al.}(2012){Ahn}, {Alexandroff}, {Allende Prieto},
  {Anderson}, {Anderton}, {Andrews}, {Aubourg}, {Bailey}, {Balbinot}, {Barnes},
  \& et~al.}]{2012ApJS..203...21A}
{Ahn}, C.~P., {Alexandroff}, R., {Allende Prieto}, C., {et~al.} 2012, \apjs,
  203, 21

\bibitem[{{Astropy Collaboration} {et~al.}(2013){Astropy Collaboration},
  {Robitaille}, {Tollerud}, {Greenfield}, {Droettboom}, {Bray}, {Aldcroft},
  {Davis}, {Ginsburg}, {Price-Whelan}, {Kerzendorf}, {Conley}, {Crighton},
  {Barbary}, {Muna}, {Ferguson}, {Grollier}, {Parikh}, {Nair}, {Unther},
  {Deil}, {Woillez}, {Conseil}, {Kramer}, {Turner}, {Singer}, {Fox}, {Weaver},
  {Zabalza}, {Edwards}, {Azalee Bostroem}, {Burke}, {Casey}, {Crawford},
  {Dencheva}, {Ely}, {Jenness}, {Labrie}, {Lim}, {Pierfederici}, {Pontzen},
  {Ptak}, {Refsdal}, {Servillat}, \& {Streicher}}]{2013A&A...558A..33A}
{Astropy Collaboration}, {Robitaille}, T.~P., {Tollerud}, E.~J., {et~al.} 2013,
  AAP, 558, A33

\bibitem[{Behroozi {et~al.}(2013{\natexlab{a}})Behroozi, Wechsler, \&
  Conroy}]{Behroozi:2012iw}
Behroozi, P.~S., Wechsler, R.~H., \& Conroy, C. 2013{\natexlab{a}}, Astrophys.
  J., 770, 57

\bibitem[{Behroozi {et~al.}(2013{\natexlab{b}})Behroozi, Wechsler, \&
  Wu}]{Behroozi:2011ju}
Behroozi, P.~S., Wechsler, R.~H., \& Wu, H.-Y. 2013{\natexlab{b}}, Astrophys.
  J., 762, 109

\bibitem[{{Belokurov} {et~al.}(2018){Belokurov}, {Erkal}, {Evans}, {Koposov},
  \& {Deason}}]{2018arXiv180203414B}
{Belokurov}, V., {Erkal}, D., {Evans}, N.~W., {Koposov}, S.~E., \& {Deason},
  A.~J. 2018, ArXiv e-prints, arXiv:1802.03414

\bibitem[{{Bland-Hawthorn} \& {Gerhard}(2016)}]{2016ARA&A..54..529B}
{Bland-Hawthorn}, J., \& {Gerhard}, O. 2016, Annual Review of Astronomy and
  Astrophysics, 54, 529

\bibitem[{{Bonaca} {et~al.}(2017){Bonaca}, {Conroy}, {Wetzel}, {Hopkins}, \&
  {Keres}}]{2017arXiv170405463B}
{Bonaca}, A., {Conroy}, C., {Wetzel}, A., {Hopkins}, P.~F., \& {Keres}, D.
  2017, ArXiv e-prints, arXiv:1704.05463

\bibitem[{{Breddels} \& {Helmi}(2013)}]{2013A&A...558L...3B}
{Breddels}, M.~A., \& {Helmi}, A. 2013, \aap, 558, L3

\bibitem[{Bullock \& Johnston(2005)}]{Bullock:2005pi}
Bullock, J.~S., \& Johnston, K.~V. 2005, Astrophys. J., 635, 931

\bibitem[{{Bullock} \& {Johnston}(2005)}]{2005ApJ...635..931B}
{Bullock}, J.~S., \& {Johnston}, K.~V. 2005, \apj, 635, 931

\bibitem[{Bullock {et~al.}(2001)Bullock, Kravtsov, \&
  Weinberg}]{Bullock:2000qf}
Bullock, J.~S., Kravtsov, A.~V., \& Weinberg, D.~H. 2001, Astrophys. J., 548,
  33

\bibitem[{{Cooper} {et~al.}(2015){Cooper}, {Parry}, {Lowing}, {Cole}, \&
  {Frenk}}]{2015MNRAS.454.3185C}
{Cooper}, A.~P., {Parry}, O.~H., {Lowing}, B., {Cole}, S., \& {Frenk}, C. 2015,
  \mnras, 454, 3185

\bibitem[{De~Lucia \& Helmi(2008)}]{DeLucia:2008gk}
De~Lucia, G., \& Helmi, A. 2008, Mon. Not. Roy. Astron. Soc., 391, 14

\bibitem[{{Deason} {et~al.}(2018){Deason}, {Belokurov}, {Koposov}, \&
  {Lancaster}}]{2018ApJ...862L...1D}
{Deason}, A.~J., {Belokurov}, V., {Koposov}, S.~E., \& {Lancaster}, L. 2018,
  \apjl, 862, L1

\bibitem[{{Deason} {et~al.}(2016){Deason}, {Mao}, \&
  {Wechsler}}]{2016ApJ...821....5D}
{Deason}, A.~J., {Mao}, Y.-Y., \& {Wechsler}, R.~H. 2016, \apj, 821, 5

\bibitem[{Diemand {et~al.}(2007)Diemand, Kuhlen, \& Madau}]{Diemand:2007qr}
Diemand, J., Kuhlen, M., \& Madau, P. 2007, Astrophys. J., 667, 859

\bibitem[{Diemand {et~al.}(2008)Diemand, Kuhlen, Madau, Zemp, Moore, Potter, \&
  Stadel}]{Diemand:2008in}
Diemand, J., Kuhlen, M., Madau, P., {et~al.} 2008, Nature, 454, 735

\bibitem[{Elahi {et~al.}(2011)Elahi, Thacker, \& Widrow}]{Elahi:2011dy}
Elahi, P.~J., Thacker, R.~J., \& Widrow, L.~M. 2011, Mon.Not.Roy.Astron.Soc.,
  418, 320

\bibitem[{{Escala} {et~al.}(2018){Escala}, {Wetzel}, {Kirby}, {Hopkins}, {Ma},
  {Wheeler}, {Kere{\v s}}, {Faucher-Gigu{\`e}re}, \& {Quataert}}]{Escala2017}
{Escala}, I., {Wetzel}, A., {Kirby}, E.~N., {et~al.} 2018, \mnras, 474, 2194

\bibitem[{Evans {et~al.}(2018)Evans, O'Hare, \& McCabe}]{Evans2018}
Evans, N.~W., O'Hare, C. A.~J., \& McCabe, C. 2018

\bibitem[{{Faucher-Gigu{\`e}re} {et~al.}(2009){Faucher-Gigu{\`e}re}, {Lidz},
  {Zaldarriaga}, \& {Hernquist}}]{2009ApJ...703.1416F}
{Faucher-Gigu{\`e}re}, C.-A., {Lidz}, A., {Zaldarriaga}, M., \& {Hernquist}, L.
  2009, \apj, 703, 1416

\bibitem[{Font {et~al.}(2006)Font, Johnston, Bullock, \&
  Robertson}]{Font:2005qs}
Font, A.~S., Johnston, K.~V., Bullock, J.~S., \& Robertson, B. 2006, Astrophys.
  J., 638, 585

\bibitem[{{Font} {et~al.}(2011){Font}, {McCarthy}, {Crain}, {Theuns}, {Schaye},
  {Wiersma}, \& {Dalla Vecchia}}]{2011MNRAS.416.2802F}
{Font}, A.~S., {McCarthy}, I.~G., {Crain}, R.~A., {et~al.} 2011, \mnras, 416,
  2802

\bibitem[{{Gaia Collaboration} {et~al.}(2018){Gaia Collaboration}, {Brown},
  {Vallenari}, {Prusti}, {de Bruijne}, {Babusiaux}, \&
  {Bailer-Jones}}]{2018arXiv180409365G}
{Gaia Collaboration}, {Brown}, A.~G.~A., {Vallenari}, A., {et~al.} 2018, ArXiv
  e-prints, arXiv:1804.09365

\bibitem[{Gallazzi {et~al.}(2005)Gallazzi, Charlot, Brinchmann, White, \&
  Tremonti}]{Gallazzi:2005df}
Gallazzi, A., Charlot, S., Brinchmann, J., White, S. D.~M., \& Tremonti, C.~A.
  2005, Mon. Not. Roy. Astron. Soc., 362, 41

\bibitem[{Garrison-Kimmel {et~al.}(2017{\natexlab{a}})Garrison-Kimmel, Bullock,
  Boylan-Kolchin, \& Bardwell}]{Garrison-Kimmel:2016szj}
Garrison-Kimmel, S., Bullock, J.~S., Boylan-Kolchin, M., \& Bardwell, E.
  2017{\natexlab{a}}, Mon. Not. Roy. Astron. Soc., 464, 3108

\bibitem[{Garrison-Kimmel
  {et~al.}(2017{\natexlab{b}})}]{Garrison-Kimmel:2017zes}
Garrison-Kimmel, S., {et~al.} 2017{\natexlab{b}}, Mon. Not. Roy. Astron. Soc.,
  471, 1709

\bibitem[{{Garrison-Kimmel} {et~al.}(2017){Garrison-Kimmel}, {Hopkins},
  {Wetzel}, {El-Badry}, {Sanderson}, {Bullock}, {Ma}, {van de Voort}, {Hafen},
  {Faucher-Gigu{\`e}re}, {Hayward}, {Quataert}, {Keres}, \&
  {Boylan-Kolchin}}]{2017arXiv171203966G}
{Garrison-Kimmel}, S., {Hopkins}, P.~F., {Wetzel}, A., {et~al.} 2017, ArXiv
  e-prints, arXiv:1712.03966

\bibitem[{{Garrison-Kimmel} {et~al.}(2018){Garrison-Kimmel}, {Hopkins},
  {Wetzel}, {Bullock}, {Boylan-Kolchin}, {Keres}, {Faucher-Giguere},
  {El-Badry}, {Lamberts}, {Quataert}, \& {Sanderson}}]{2018arXiv180604143G}
---. 2018, ArXiv e-prints, arXiv:1806.04143

\bibitem[{{G{\'o}mez} {et~al.}(2010){G{\'o}mez}, {Helmi}, {Brown}, \&
  {Li}}]{2010MNRAS.408..935G}
{G{\'o}mez}, F.~A., {Helmi}, A., {Brown}, A. G.~A., \& {Li}, Y.-S. 2010,
  \mnras, 408, 935

\bibitem[{{Grillmair} \& {Carlin}(2016)}]{2016ASSL..420...87G}
{Grillmair}, C.~J., \& {Carlin}, J.~L. 2016, in Astrophysics and Space Science
  Library, Vol. 420, Tidal Streams in the Local Group and Beyond, ed. H.~J.
  {Newberg} \& J.~L. {Carlin}, 87

\bibitem[{Guedes {et~al.}(2011)Guedes, Callegari, Madau, \&
  Mayer}]{Guedes:2011ux}
Guedes, J., Callegari, S., Madau, P., \& Mayer, L. 2011, Astrophys. J., 742, 76

\bibitem[{{Helmi} {et~al.}(2018){Helmi}, {Babusiaux}, {Koppelman}, {Massari},
  {Veljanoski}, \& {Brown}}]{2018arXiv180606038H}
{Helmi}, A., {Babusiaux}, C., {Koppelman}, H.~H., {et~al.} 2018, ArXiv
  e-prints, arXiv:1806.06038

\bibitem[{Helmi \& White(1999)}]{Helmi:1999ks}
Helmi, A., \& White, S. D.~M. 1999, Mon. Not. Roy. Astron. Soc., 307, 495

\bibitem[{{Helmi} {et~al.}(1999){Helmi}, {White}, {de Zeeuw}, \&
  {Zhao}}]{1999Natur.402...53H}
{Helmi}, A., {White}, S.~D.~M., {de Zeeuw}, P.~T., \& {Zhao}, H. 1999, \nat,
  402, 53

\bibitem[{Helmi {et~al.}(2003)Helmi, White, \& Springel}]{Helmi:2002iu}
Helmi, A., White, S. D.~M., \& Springel, V. 2003, Mon. Not. Roy. Astron. Soc.,
  339, 834

\bibitem[{Herzog-Arbeitman {et~al.}(2018{\natexlab{a}})Herzog-Arbeitman,
  Lisanti, Madau, \& Necib}]{Herzog-Arbeitman:2017fte}
Herzog-Arbeitman, J., Lisanti, M., Madau, P., \& Necib, L. 2018{\natexlab{a}},
  Phys. Rev. Lett., 120, 041102

\bibitem[{Herzog-Arbeitman {et~al.}(2018{\natexlab{b}})Herzog-Arbeitman,
  Lisanti, \& Necib}]{Herzog-Arbeitman:2017zbm}
Herzog-Arbeitman, J., Lisanti, M., \& Necib, L. 2018{\natexlab{b}}, JCAP, 1804,
  052

\bibitem[{Hopkins(2015)}]{Hopkins:2014qka}
Hopkins, P.~F. 2015, Mon. Not. Roy. Astron. Soc., 450, 53

\bibitem[{Hopkins {et~al.}(2013)Hopkins, Narayanan, \&
  Murray}]{Hopkins:2013oba}
Hopkins, P.~F., Narayanan, D., \& Murray, N. 2013, Mon. Not. Roy. Astron. Soc.,
  4, 432

\bibitem[{{Hopkins} {et~al.}(2018){Hopkins}, {Wetzel}, {Kere{\v{s}}},
  {Faucher-Gigu{\`e}re}, {Quataert}, {Boylan-Kolchin}, {Murray}, {Hayward},
  {Garrison-Kimmel}, {Hummels}, {Feldmann}, {Torrey}, {Ma},
  {Angl{\'e}s-Alc{\'a}zar}, {Su}, {Orr}, {Schmitz}, {Escala}, {Sanderson},
  {Grudi{\'c}}, {Hafen}, {Kim}, {Fitts}, {Bullock}, {Wheeler}, {Chan},
  {Elbert}, \& {Narayanan}}]{2017arXiv170206148H}
{Hopkins}, P.~F., {Wetzel}, A., {Kere{\v{s}}}, D., {et~al.} 2018, \mnras, 480,
  800

\bibitem[{Ivezic {et~al.}(2000)}]{Ivezic:2000ua}
Ivezic, Z., {et~al.} 2000, Astron.J., 120, 963

\bibitem[{Johnston {et~al.}(1996)Johnston, Hernquist, \&
  Bolte}]{Johnston:1996sb}
Johnston, K.~V., Hernquist, L., \& Bolte, M. 1996, Astrophys. J., 465, 278

\bibitem[{Johnston {et~al.}(1995)Johnston, Spergel, \&
  Hernquist}]{Johnston:1995vd}
Johnston, K.~V., Spergel, D.~N., \& Hernquist, L. 1995, Astrophys. J., 451, 598

\bibitem[{Katz \& White(1993)}]{KatzWhite1993}
Katz, N., \& White, S. D.~M. 1993, Astrophys. J., 412, 455

\bibitem[{Kirby {et~al.}(2013)Kirby, Cohen, Guhathakurta, Cheng, Bullock, \&
  Gallazzi}]{Kirby:2013wna}
Kirby, E.~N., Cohen, J.~G., Guhathakurta, P., {et~al.} 2013, Astrophys. J.,
  779, 102

\bibitem[{{Klypin} {et~al.}(2011){Klypin}, {Trujillo-Gomez}, \&
  {Primack}}]{2011ApJ...740..102K}
{Klypin}, A.~A., {Trujillo-Gomez}, S., \& {Primack}, J. 2011, \apj, 740, 102

\bibitem[{Kroupa(2001)}]{Kroupa:2000iv}
Kroupa, P. 2001, Mon. Not. Roy. Astron. Soc., 322, 231

\bibitem[{Krumholz \& Gnedin(2011)}]{Krumholz:2010wm}
Krumholz, M.~R., \& Gnedin, N.~Y. 2011, Astrophys. J., 729, 36

\bibitem[{Kuhlen {et~al.}(2012)Kuhlen, Lisanti, \& Spergel}]{Kuhlen:2012fz}
Kuhlen, M., Lisanti, M., \& Spergel, D.~N. 2012, Phys. Rev., D86, 063505

\bibitem[{Kuhlen {et~al.}(2010)Kuhlen, Weiner, Diemand, Madau, Moore, Potter,
  Stadel, \& Zemp}]{Kuhlen:2009vh}
Kuhlen, M., Weiner, N., Diemand, J., {et~al.} 2010, JCAP, 1002, 030

\bibitem[{{Leitherer} {et~al.}(2014){Leitherer}, {Ekstr{\"o}m}, {Meynet},
  {Schaerer}, {Agienko}, \& {Levesque}}]{2014ApJS..212...14L}
{Leitherer}, C., {Ekstr{\"o}m}, S., {Meynet}, G., {et~al.} 2014, \apjs, 212, 14

\bibitem[{{Leitherer} {et~al.}(1999){Leitherer}, {Schaerer}, {Goldader},
  {Delgado}, {Robert}, {Kune}, {de Mello}, {Devost}, \&
  {Heckman}}]{1999ApJS..123....3L}
{Leitherer}, C., {Schaerer}, D., {Goldader}, J.~D., {et~al.} 1999, \apjs, 123,
  3

\bibitem[{{Lindegren} {et~al.}(2016){Lindegren}, {Lammers}, {Bastian},
  {Hern{\'a}ndez}, {Klioner}, {Hobbs}, {Bombrun}, {Michalik}, {Ramos-Lerate},
  {Butkevich}, {Comoretto}, {Joliet}, {Holl}, {Hutton}, {Parsons},
  {Steidelm{\"u}ller}, {Abbas}, {Altmann}, {Andrei}, {Anton}, {Bach},
  {Barache}, {Becciani}, {Berthier}, {Bianchi}, {Biermann}, {Bouquillon},
  {Bourda}, {Br{\"u}semeister}, {Bucciarelli}, {Busonero}, {Carlucci},
  {Casta{\~n}eda}, {Charlot}, {Clotet}, {Crosta}, {Davidson}, {de Felice},
  {Drimmel}, {Fabricius}, {Fienga}, {Figueras}, {Fraile}, {Gai}, {Garralda},
  {Geyer}, {Gonz{\'a}lez-Vidal}, {Guerra}, {Hambly}, {Hauser}, {Jordan},
  {Lattanzi}, {Lenhardt}, {Liao}, {L{\"o}ffler}, {McMillan}, {Mignard}, {Mora},
  {Morbidelli}, {Portell}, {Riva}, {Sarasso}, {Serraller}, {Siddiqui}, {Smart},
  {Spagna}, {Stampa}, {Steele}, {Taris}, {Torra}, {van Reeven}, {Vecchiato},
  {Zschocke}, {de Bruijne}, {Gracia}, {Raison}, {Lister}, {Marchant},
  {Messineo}, {Soffel}, {Osorio}, {de Torres}, \&
  {O'Mullane}}]{2016A&A...595A...4L}
{Lindegren}, L., {Lammers}, U., {Bastian}, U., {et~al.} 2016, \aap, 595, A4

\bibitem[{Lisanti \& Spergel(2012)}]{Lisanti:2011as}
Lisanti, M., \& Spergel, D.~N. 2012, Phys. Dark Univ., 1, 155

\bibitem[{Lisanti {et~al.}(2015)Lisanti, Spergel, \& Madau}]{Lisanti:2014dva}
Lisanti, M., Spergel, D.~N., \& Madau, P. 2015, Astrophys. J., 807, 14

\bibitem[{Ma {et~al.}(2016)Ma, Hopkins, Faucher-Giguere, Zolman, Muratov,
  Keres, \& Quataert}]{Ma:2015ota}
Ma, X., Hopkins, P.~F., Faucher-Giguere, C.-A., {et~al.} 2016, Mon. Not. Roy.
  Astron. Soc., 456, 2140

\bibitem[{Ma {et~al.}(2017)Ma, Hopkins, Wetzel, Kirby, Angles-Alcazar,
  Faucher-Giguere, Keres, \& Quataert}]{Ma:2016fbd}
Ma, X., Hopkins, P.~F., Wetzel, A.~R., {et~al.} 2017, Mon. Not. Roy. Astron.
  Soc., 467, 2430

\bibitem[{Maciejewski {et~al.}(2011)Maciejewski, Vogelsberger, White, \&
  Springel}]{Maciejewski:2010gz}
Maciejewski, M., Vogelsberger, M., White, S. D.~M., \& Springel, V. 2011, Mon.
  Not. Roy. Astron. Soc., 415, 2475

\bibitem[{{McCarthy} {et~al.}(2012){McCarthy}, {Font}, {Crain}, {Deason},
  {Schaye}, \& {Theuns}}]{2012MNRAS.420.2245M}
{McCarthy}, I.~G., {Font}, A.~S., {Crain}, R.~A., {et~al.} 2012, \mnras, 420,
  2245

\bibitem[{{McConnachie}(2012)}]{2012AJ....144....4M}
{McConnachie}, A.~W. 2012, \aj, 144, 4

\bibitem[{Myeong {et~al.}(2018{\natexlab{a}})Myeong, Evans, Belokurov,
  Amorisco, \& Koposov}]{Myeong:2017skt}
Myeong, G.~C., Evans, N.~W., Belokurov, V., Amorisco, N.~C., \& Koposov, S.
  2018{\natexlab{a}}, Mon. Not. Roy. Astron. Soc., 475, 1537

\bibitem[{Myeong {et~al.}(2018{\natexlab{b}})Myeong, Evans, Belokurov, Sanders,
  \& Koposov}]{Myeong:2018kfh}
Myeong, G.~C., Evans, N.~W., Belokurov, V., Sanders, J.~L., \& Koposov, S.~E.
  2018{\natexlab{b}}, arXiv:1805.00453

\bibitem[{{Myeong} {et~al.}(2018){Myeong}, {Evans}, {Belokurov}, {Sanders}, \&
  {Koposov}}]{2018arXiv180407050M}
{Myeong}, G.~C., {Evans}, N.~W., {Belokurov}, V., {Sanders}, J.~L., \&
  {Koposov}, S.~E. 2018, ArXiv e-prints, arXiv:1804.07050

\bibitem[{Necib {et~al.}(2018)Necib, Lisanti, \& Belokurov}]{necib2018}
Necib, L., Lisanti, M., \& Belokurov, V. 2018, arXiv:1807.02519

\bibitem[{O'Hare {et~al.}(2018)O'Hare, McCabe, Evans, Myeong, \&
  Belokurov}]{OHare:2018trr}
O'Hare, C. A.~J., McCabe, C., Evans, N.~W., Myeong, G., \& Belokurov, V. 2018,
  arXiv:1807.09004

\bibitem[{Onorbe {et~al.}(2014)Onorbe, Garrison-Kimmel, Maller, Bullock, Rocha,
  \& Hahn}]{Onorbe2014}
Onorbe, J., Garrison-Kimmel, S., Maller, A.~H., {et~al.} 2014, Mon. Not. Roy.
  Astron. Soc., 437, 1894

\bibitem[{{Pe{\~n}arrubia} {et~al.}(2008){Pe{\~n}arrubia}, {Navarro}, \&
  {McConnachie}}]{2008AN....329..934P}
{Pe{\~n}arrubia}, J., {Navarro}, J.~F., \& {McConnachie}, A.~W. 2008,
  Astronomische Nachrichten, 329, 934

\bibitem[{P\'erez \& Granger(2007)}]{PER-GRA:2007}
P\'erez, F., \& Granger, B.~E. 2007, Computing in Science and Engineering, 9,
  21

\bibitem[{Pillepich {et~al.}(2015)Pillepich, Madau, \&
  Mayer}]{Pillepich:2014jfa}
Pillepich, A., Madau, P., \& Mayer, L. 2015, Astrophys. J., 799, 184

\bibitem[{Purcell {et~al.}(2007)Purcell, Bullock, \& Zentner}]{Purcell:2007tr}
Purcell, C.~W., Bullock, J.~S., \& Zentner, A.~R. 2007, Astrophys. J., 666, 20

\bibitem[{{Purcell} {et~al.}(2012){Purcell}, {Zentner}, \&
  {Wang}}]{2012JCAP...08..027P}
{Purcell}, C.~W., {Zentner}, A.~R., \& {Wang}, M.-Y. 2012, \jcap, 8, 027

\bibitem[{Robertson {et~al.}(2005)Robertson, Bullock, Font, Johnston, \&
  Hernquist}]{Robertson:2005gv}
Robertson, B., Bullock, J.~S., Font, A.~S., Johnston, K.~V., \& Hernquist, L.
  2005, Astrophys. J., 632, 872

\bibitem[{{Sanderson} {et~al.}(2017){Sanderson}, {Garrison-Kimmel}, {Wetzel},
  {Keung Chan}, {Hopkins}, {Kere{\v s}}, {Escala}, {Faucher-Gigu{\`e}re}, \&
  {Ma}}]{2017arXiv171205808S}
{Sanderson}, R.~E., {Garrison-Kimmel}, S., {Wetzel}, A., {et~al.} 2017, ArXiv
  e-prints, arXiv:1712.05808

\bibitem[{{Sanderson} {et~al.}(2018){Sanderson}, {Wetzel}, {Loebman}, {Sharma},
  {Hopkins}, {Garrison-Kimmel}, {Faucher-Gigu{\`e}re}, {Kere{\v{s}}}, \&
  {Quataert}}]{2018arXiv180610564S}
{Sanderson}, R.~E., {Wetzel}, A., {Loebman}, S., {et~al.} 2018, ArXiv e-prints,
  arXiv:1806.10564

\bibitem[{Springel(2005)}]{Springel:2005mi}
Springel, V. 2005, Mon. Not. Roy. Astron. Soc., 364, 1105

\bibitem[{Springel {et~al.}(2008)}]{Springel:2008cc}
Springel, V., {et~al.} 2008, Mon. Not. Roy. Astron. Soc., 391, 1685

\bibitem[{{Su} {et~al.}(2017){Su}, {Hopkins}, {Hayward}, {Faucher-Gigu{\`e}re},
  {Kere{\v{s}}}, {Ma}, \& {Robles}}]{2017MNRAS.471..144S}
{Su}, K.-Y., {Hopkins}, P.~F., {Hayward}, C.~C., {et~al.} 2017, \mnras, 471,
  144

\bibitem[{{Vogelsberger} \& {White}(2011)}]{2011MNRAS.413.1419V}
{Vogelsberger}, M., \& {White}, S.~D.~M. 2011, \mnras, 413, 1419

\bibitem[{Vogelsberger {et~al.}(2009)Vogelsberger, Helmi, Springel, White,
  Wang, Frenk, Jenkins, Ludlow, \& Navarro}]{Vogelsberger:2008qb}
Vogelsberger, M., Helmi, A., Springel, V., {et~al.} 2009, Mon. Not. Roy.
  Astron. Soc., 395, 797

\bibitem[{{Wang} {et~al.}(2011){Wang}, {Navarro}, {Frenk}, {White}, {Springel},
  {Jenkins}, {Helmi}, {Ludlow}, \& {Vogelsberger}}]{2011MNRAS.413.1373W}
{Wang}, J., {Navarro}, J.~F., {Frenk}, C.~S., {et~al.} 2011, \mnras, 413, 1373

\bibitem[{Wetzel {et~al.}(2016)Wetzel, Hopkins, Kim, Faucher-Giguere, Keres, \&
  Quataert}]{Wetzel2016}
Wetzel, A.~R., Hopkins, P.~F., Kim, J.-h., {et~al.} 2016, Astrophys. J., 827,
  L23

\bibitem[{{Wetzel} {et~al.}(2016){Wetzel}, {Hopkins}, {Kim},
  {Faucher-Gigu{\`e}re}, {Kere{\v s}}, \& {Quataert}}]{2016ApJ...827L..23W}
{Wetzel}, A.~R., {Hopkins}, P.~F., {Kim}, J.-h., {et~al.} 2016, \apjl, 827, L23

\bibitem[{{Wheeler} {et~al.}(2015){Wheeler}, {O{\~n}orbe}, {Bullock},
  {Boylan-Kolchin}, {Elbert}, {Garrison- Kimmel}, {Hopkins}, \&
  {Kere{\v{s}}}}]{2015MNRAS.453.1305W}
{Wheeler}, C., {O{\~n}orbe}, J., {Bullock}, J.~S., {et~al.} 2015, \mnras, 453,
  1305

\bibitem[{{White} \& {Rees}(1978)}]{1978MNRAS.183..341W}
{White}, S.~D.~M., \& {Rees}, M.~J. 1978, \mnras, 183, 341

\bibitem[{Yanny {et~al.}(2000)}]{Yanny:2000ty}
Yanny, B., {et~al.} 2000, Astrophys.J., 540, 825

\bibitem[{Zemp {et~al.}(2009)Zemp, Diemand, Kuhlen, Madau, Moore,
  {et~al.}}]{Zemp:2008gw}
Zemp, M., Diemand, J., Kuhlen, M., {et~al.} 2009, Mon. Not. Roy. Astron. Soc.,
  394, 641

\bibitem[{{Zolotov} {et~al.}(2009){Zolotov}, {Willman}, {Brooks}, {Governato},
  {Brook}, {Hogg}, {Quinn}, \& {Stinson}}]{2009ApJ...702.1058Z}
{Zolotov}, A., {Willman}, B., {Brooks}, A.~M., {et~al.} 2009, \apj, 702, 1058

\end{thebibliography}

\newpage

\onecolumngrid

\newpage
\appendix

\setcounter{equation}{0}
\setcounter{figure}{0}
\setcounter{table}{0}
\setcounter{section}{0}
\makeatletter
\renewcommand{\theequation}{S\arabic{equation}}
\renewcommand{\thefigure}{S\arabic{figure}}
\renewcommand{\thetable}{S\arabic{table}}

\begin{figure*}[h] 
   \centering
   	\includegraphics[width=0.95\textwidth]{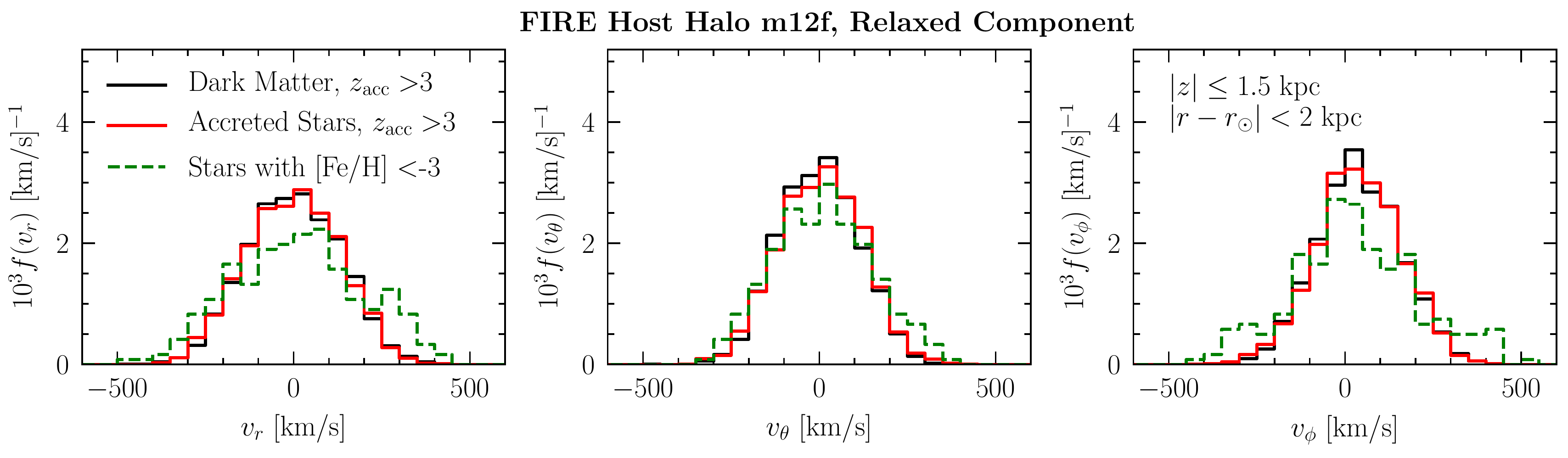} 
	\includegraphics[width=0.95\textwidth]{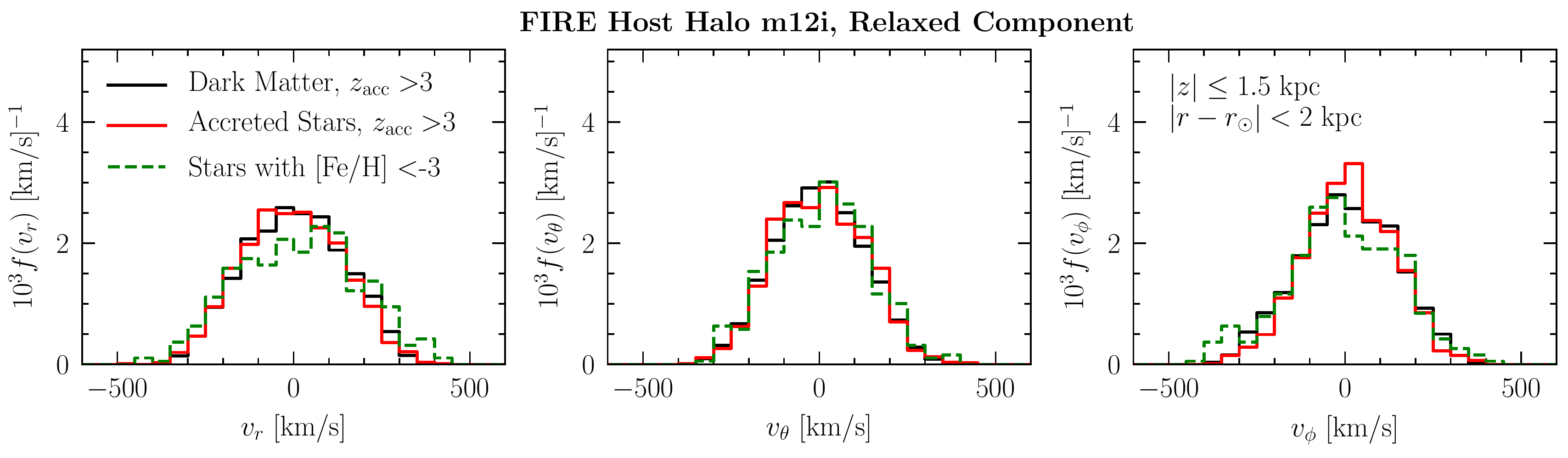} 
   \caption{As in \Fig{fig:virialized}, this figure shows the present-day distributions for the stars (red solid) and dark matter (black solid) accreted before redshift $z_{\rm{acc}} > 3$ in \mf~(top) and \mi~(bottom).  Here, however, we show the corresponding distributions for all stars (not just the  accreted subset) with $\FeH <-3$ (green dashed), as opposed to $\FeH <-2$.}
   \label{fig:virialized_m3}
\end{figure*}

\begin{figure*}[h] 
   \centering
	\includegraphics[width=7in]{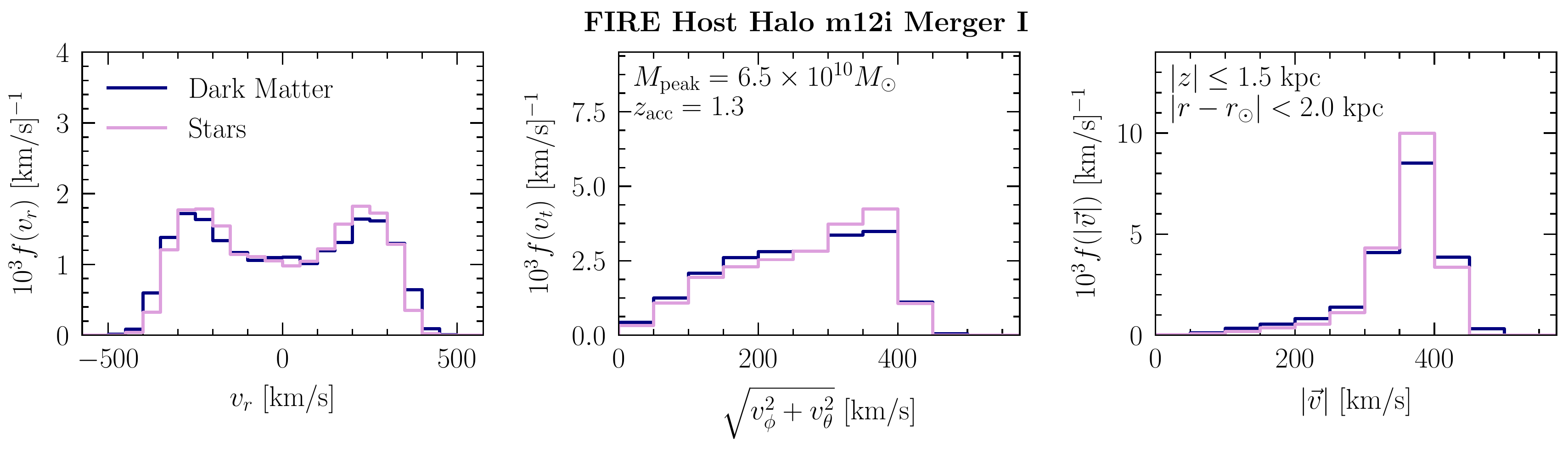} 
	\includegraphics[width=7in]{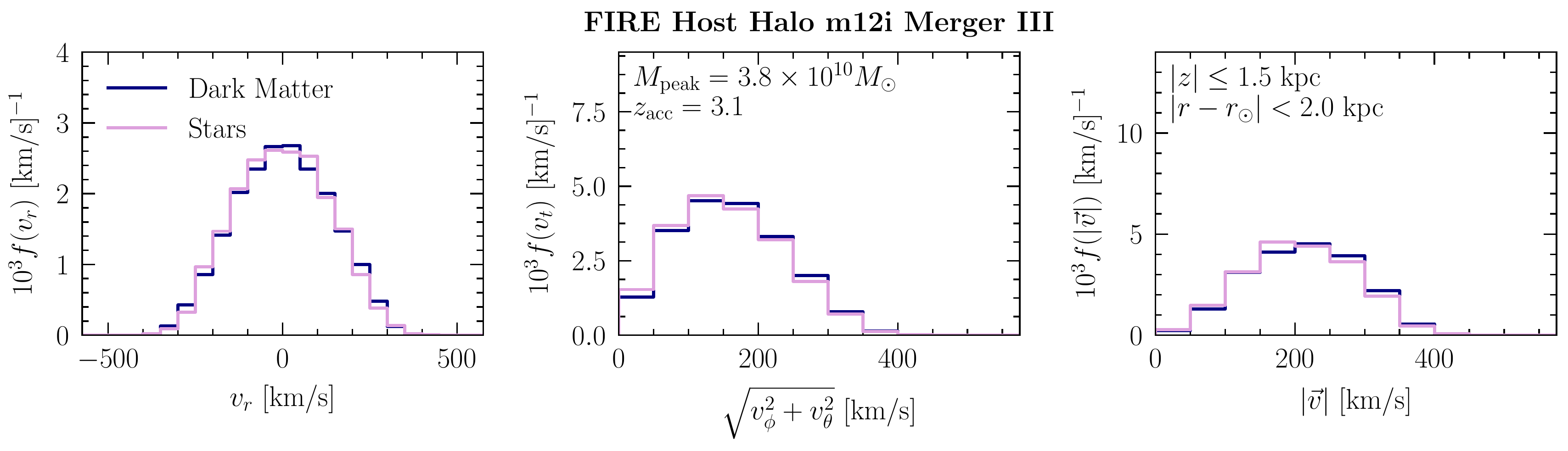} 
   \caption{Same as \Fig{fig:streamdebris}, except for Merger~I and III of \mi. }
   \label{fig:other_mergers}
\end{figure*}

\begin{figure*}[h] 
   \centering
	\includegraphics[width=7in]{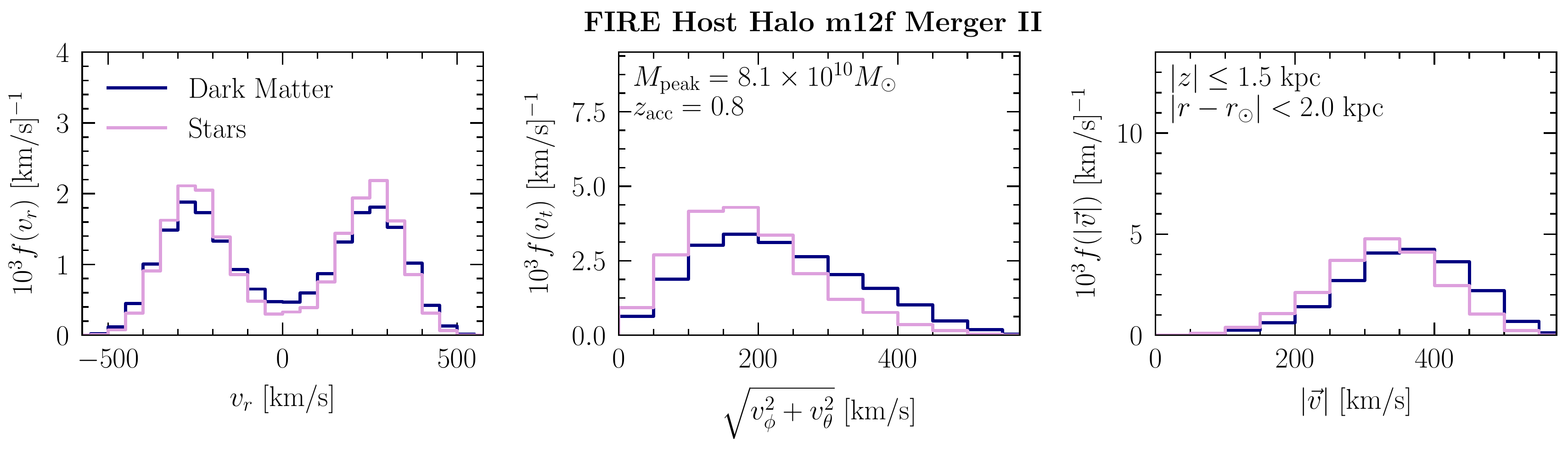} 
	\includegraphics[width=7in]{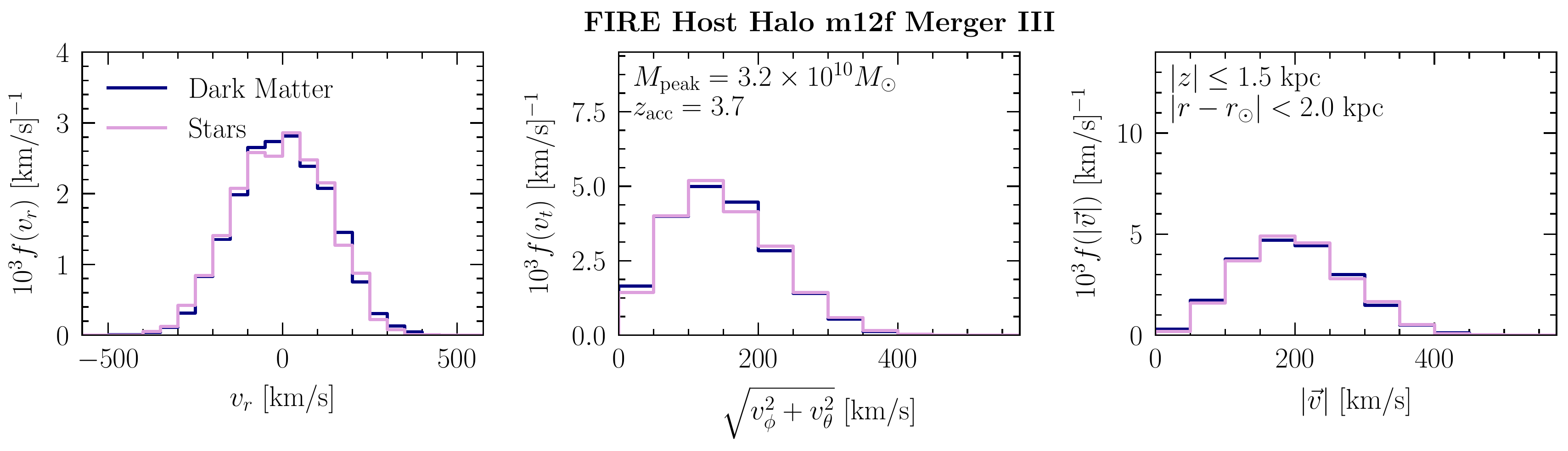} 
   \caption{Same as \Fig{fig:streamdebris}, except for Merger~II and III of \mf. }
   \label{fig:other_mergers_mf}
\end{figure*}

\begin{figure*}[h] 
   \centering
	\includegraphics[width=0.95\textwidth]{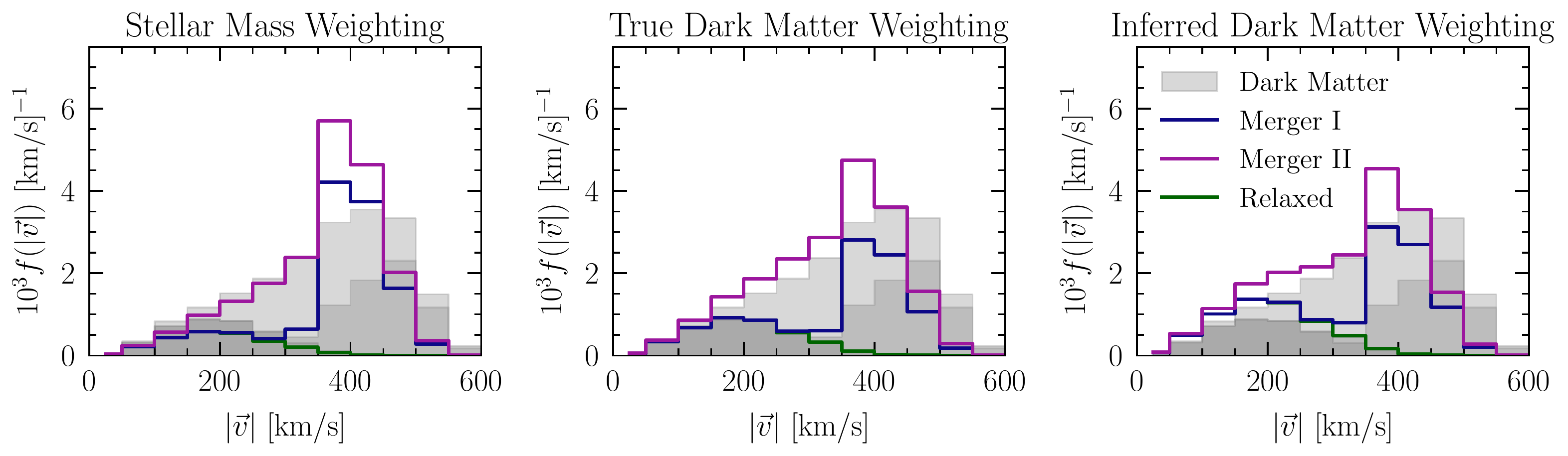}
   \caption{Same as \Fig{fig:scaling_relation}, except for \mf. 
}
   \label{fig:scaling_relation_f}
\end{figure*}

\begin{figure*}[h] 
   \centering
	\includegraphics[width=0.95\textwidth]{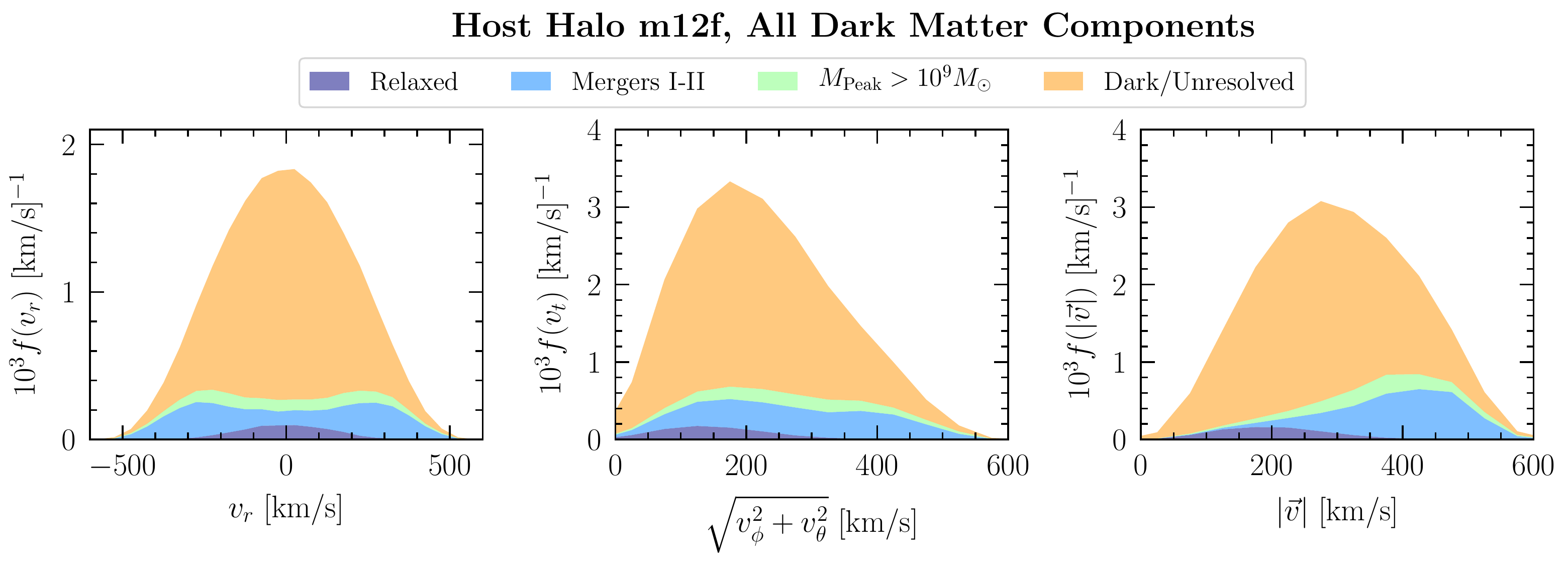}
   \caption{ Same as \Fig{fig:unvirializedmi}, except for \mf.  Comparing Figs. \ref{fig:unvirializedmi} and \ref{fig:unvirializedmf} to a Maxwell-Boltzmann distribution with a dispersion $220/\sqrt{2}$ km/s, we find that the \textsc{Fire} distributions are peaked at larger speeds of $\sim 275$ km/s than the Standard Halo Model, which is peaked at $220$~km/s.}
   \label{fig:unvirializedmf}
\end{figure*}

In this Appendix, we provide some additional figures that supplement the discussion in the main text.  \Fig{fig:virialized_m3} compares the velocity distributions of the relaxed populations to that of stars with $\FeH < -3$.  Figs.~\ref{fig:other_mergers} and \ref{fig:other_mergers_mf} shows the velocity distributions of other significant mergers in \mi~and \mf.  \Fig{fig:scaling_relation_f} shows the results of estimating the total dark matter distribution in \mf, and \Fig{fig:unvirializedmf} plots the velocity distributions for all the dark matter components in \mf.
\vspace{1in}

\clearpage

\clearpage

\vspace{0.5in}

\end{document}